\documentclass[amsmath,amssymb,aps,pra,reprint,groupedaddress,showpacs,superscriptaddress]{revtex4-1}

\usepackage{graphicx}
\usepackage{dcolumn}
\usepackage{bm}
\usepackage{longtable}

\newcommand{\fq}{\mathbb{F}_{q}}
\newcommand{\tr}{^{T}}
\newcommand{\rk}[1]{\textup{\mbox{\,rank}}\,#1}

\newcommand{\qecci}[4]{\left[\left[ #1, #2, #3 ; #4 \right]\right]}
\newcommand{\Proof}{\noindent\textbf{Proof.}\quad}
\newcommand{\qed}{\hfill$\Box$}
\newcommand{\supp}{\mbox{supp}}

\newcommand{\ka}{\mu}

\newtheorem{theorem}{Theorem}
\newtheorem{lemma}[theorem]{Lemma}

\newtheorem{proposition}[theorem]{Proposition}

\begin{document}

\title{Entanglement-assisted quantum low-density parity-check codes}

\author{Yuichiro Fujiwara}
\email[]{yfujiwar@mtu.edu}
\altaffiliation[Also at ]{Graduate School of System and Information Engineering, University of Tsukuba.}

\author{David Clark}
\email[]{dcclark@mtu.edu}
\affiliation{Department of Mathematical Sciences, Michigan Technological University, Houghton,
MI 49931 USA}

\author{Peter Vandendriessche}
\author{Maarten De Boeck}
\affiliation{Department of Mathematics, Ghent University, Krijgslaan 281-S22, 9000 Ghent, Belgium}

\author{Vladimir D. Tonchev}
\affiliation{Department of Mathematical Sciences, Michigan Technological University, Houghton,
MI 49931 USA}

\date{\today}

\begin{abstract}
This paper develops a general method for constructing entanglement-assisted quantum 
low-density parity-check (LDPC) codes, which is based on combinatorial design theory. 
Explicit constructions are given for entanglement-assisted quantum error-correcting 
codes (EAQECCs) with many desirable properties. These properties include the 
requirement of only one initial entanglement bit, high error correction performance, 
high rates, and low decoding complexity.
The proposed method produces several infinite families of new codes
with a wide variety of parameters and entanglement requirements.
Our framework encompasses the previously known entanglement-assisted quantum LDPC codes
having the best error correction performance
and many new codes with better block error rates in simulations over the depolarizing channel.
We also determine important parameters of several well-known classes of quantum and classical LDPC codes for previously unsettled cases.
\end{abstract}

\pacs{03.67.Hk, 03.67.Mn, 03.67.Pp}

\maketitle

\section{\label{intro}Introduction}
This paper develops a general combinatorial method for constructing quantum low-density parity-check (LDPC) codes
under the entanglement-assisted stabilizer formalism established by Brun, Devetak, and Hsieh \cite{BDH}.
Our results include many new explicit constructions for entanglement-assisted quantum error-correcting codes for a wide range of parameters.
We also prove a variety of new results for classical error-correcting codes, which directly apply to the quantum setting.
Most of the quantum codes designed in this paper achieve high error correction performance and high rates while requiring prescribed amounts of entanglement.
These codes can be efficiently decoded by message-passing algorithms such as the sum-product algorithm (for details of iterative probabilistic decoding, see \cite{Mbook}).

The existence of quantum error-correcting codes was one of the most important discoveries in quantum information science \cite{Shor, Steane}.
Unfortunately, most of the known quantum error-correcting codes lack practical decoding algorithms.

In this paper, we focus on the use of LDPC codes in a quantum setting.
Classical LDPC codes \cite{Gallager} can be efficiently decoded
while achieving information rates close to the classical Shannon limit \cite{LMSS, RU, RSU}.
This extends to the quantum setting: the pioneering works of Hagiwara and Imai \cite{Hagiwara} and MacKay, Mitchison, and McFadden \cite{MMM}
presented quantum LDPC codes which surpassed, in simulations,
all previously known quantum error-correcting codes.
Their quantum codes have nearly as low decoding complexity as their classical counterparts.

However, most of the previous results concerning quantum LDPC codes and related efficiently decodable codes have relied on the stabilizer formalism,
which severely restricts the classical codes which can be used.
The difficulty in developing constructions for non-stabilizer codes was also a substantial obstacle.

Our results use the newly developed theory of entanglement-assisted quantum error-correcting codes (EAQECCs) \cite{Bowen, BDH, Catalytic, DBH}.
The entanglement-assisted stabilizer formalism allows the use of arbitrary classical binary or quaternary linear codes
for quantum data transmission and error correction by using shared entanglement \cite{HDB, WB}.
Previous work related to entanglement-assisted quantum LDPC codes is due to Hsieh, Brun, and Devetak \cite{HBD} and Hsieh, Yen, and Hsu \cite{HYH}.

The major difficulty in using classical LDPC codes in the entanglement-assisted quantum setting is that
very little is known about methods for designing EAQECCs requiring desirable amounts of entanglement.
While entanglement-assisted quantum LDPC codes can achieve both notable error correction performance and low decoding complexity,
the resulting quantum codes might require too much entanglement to be usable; in general entanglement is a valuable resource \cite{WB}.
In some situations, one might wish to effectively take advantage of high performance codes requiring a larger amount of entanglement \cite{Catalytic, BDH}.
To the best of the authors' knowledge, no general methods have been developed which allow the code designer flexibility in choice of parameters and required amounts of entanglement. 

Our primary focus in this paper is to show that it is possible to create infinite classes of EAQECCs
which consume prescribed amounts of entanglement and achieve good error correction performance while allowing efficient decoding.
Our methods are flexible and address various situations, including the extreme case when an EAQECC requires only one preexisting entanglement bit.

The entanglement-assisted quantum LDPC codes which we construct include
quantum analogues of the well-known finite geometry LDPC codes
originally proposed by Kou, Lin, and Fossorier \cite{KLF} (see also \cite{TXKLA, TXLA}),
and LDPC codes from balanced incomplete block designs that achieve the upper bound on the rate for a classical regular LDPC code with girth six
proposed independently by several authors (see \cite{JohnsonBook} and references therein).
Some classes of our codes outperform previously proposed quantum LDPC codes having the best known error correction performance \cite{Hagiwara,MMM,HBD,HYH}.

Our primary tools come from combinatorial design theory, which plays an important role in classical coding theory \cite{T-h}
and also gave several classes of stabilizer codes in quantum coding theory \cite{Aly, Djordjevic, Djordjevic4, T-cap, T-gm}.
The use of combinatorial design theory allows us to exactly determine or give tighter bounds
on the parameters of the finite geometry LDPC codes in both quantum and classical settings.
Comprehensive lists of the parameters of these codes are given
in Tables \ref{fig:quantumsummary} and \ref{fig:classicalsummary} in Appendix \ref{appendix2}.

In Section \ref{LDPC_designs}, we outline our framework
for designing entanglement-assisted quantum LDPC codes by using combinatorial design theory.
Section \ref{finite} gives explicit constructions for entanglement-assisted quantum LDPC codes based on finite geometries and related combinatorial structures.
New results concerning the well-known classical finite geometry LDPC codes are also given in this section.
Section \ref{sim} presents simulation results of our entanglement-assisted quantum LDPC codes and discusses their performance over the depolarizing channel.
Section \ref{conclude} contains concluding remarks and discusses some related
problems that can be treated with the techniques developed in this paper.

\section{\label{LDPC_designs}Combinatorial entanglement-assisted quantum LDPC codes}
In this section we give a general construction method for entanglement-assisted quantum LDPC codes based on combinatorial designs.
We do not describe the theory of classical LDPC codes in detail here, instead referring the reader to \cite{Mbook, JohnsonBook} and references therein.
Relations between quantum error-correcting codes and LDPC codes are concisely yet thoroughly explained in \cite{MMM, HBD}.
Basic notions related to LDPC codes and their relations to combinatorial designs can be found in \cite{AHKXL}.
For a detailed treatment of the entanglement-assisted stabilizer formalism, we refer the reader to \cite{BDH, Catalytic, DBH, HDB}.

In Subsection \ref{preliminaries} we introduce necessary notions from coding theory and combinatorial design theory.
A general method for designing entanglement-assisted quantum LDPC codes is presented in Subsection \ref{principles}.

\subsection{\label{preliminaries}Preliminaries}
An $[[n,k;c]]$ \textit{entanglement-assisted quantum error-correcting code} (EAQECC) encodes $k$ logical qubits
into $n$ physical qubits with the help of $c$ copies of maximally entangled states.
As in classical coding theory, $n$ is the \textit{length} of the EAQECC, and $k$ the \textit{dimension}.
We say that the EAQECC requires $c$ \textit{ebits}.
An $[[n,k;c]]$ EAQECC with \textit{distance} $d$ will be referred to as an $[[n,k,d;c]]$ code.

The \textit{rate} of an $[[n,k;c]]$ EAQECC is defined to be $\frac{k}{n}$.
The ratio $\frac{k-c}{n}$ is called the \textit{net rate}.
The latter figure describes the rate of an EAQECC
when used as a catalytic quantum error-correcting codes to create $c$ new bits of shared entanglement \cite{BDH, Catalytic}.

Throughout this paper, matrix operations are performed over ${\mathbb{F}_{2}}$, the finite field of order two.
The ranks of matrices are also calculated over ${\mathbb{F}_{2}}$.

We employ the Calderbank-Shor-Steane (CSS) construction \cite{CS, Steane, BDH, HDB}.
Usually the CSS construction uses a minimal set of independent generators to construct an EAQECC.
Hence, the construction is often described by using a classical binary linear code with a parity-check matrix of full rank.
However, in actual decoding steps, sparse-graph codes may take advantage of redundant parity-check equations
to improve error correction performance. Because the extended syndrome can be obtained in polynomial time
without additional quantum interactions, we use the following formulation of the CSS construction for EAQECCs.
\begin{theorem}[Hsieh, Brun, and Devetak \cite{HBD}]\label{thm:css}
If there exists a classical binary $[n,k,d]$ code with parity-check matrix $H$, then there exists an $[[n,2k-n+c,d;c]]$ \textup{EAQECC}, where $c = \rk{HH^T}$.
\end{theorem}
Note that $H$ may contain redundant rows which are related only to classical operations to infer the noise by a message-passing algorithm.

We apply Theorem \ref{thm:css} to classical sparse-graph codes.
An LDPC code is typically defined as a binary linear code with parity-check matrix $H$ in which every row and column is sparse.
In this paper we consider LDPC codes with parity-check matrices whose rows and columns contain only small numbers of ones
so that simple message-passing algorithms can efficiently give good performance in decoding.
\begin{proposition}
An \textup{LDPC} code with parity-check matrix $H$ with $n$ columns and minimum distance $d$
defines a classical binary $[n,n-\rk{H},d]$ code, which yields an $[[n, n - 2 \rk{H} + \rk{HH^T}, d; \rk{HH^T}]]$ \textup{EAQECC}.
\end{proposition}

The \textit{Tanner graph} of an $m \times n$ parity-check matrix $H$ is the bipartite graph consisting of $n$ bit vertices and $m$ parity-check vertices,
where an edge joins a bit vertex to a parity-check vertex if that bit is included in the corresponding parity-check equation.
A \textit{cycle} in a graph is a sequence of connected vertices which starts and ends at the same vertex in the graph and contains no other vertices more than once.
The \textit{girth} of a parity-check matrix is the length of a shortest cycle in the corresponding Tanner graph.
Short cycles can severely reduce the performance of an otherwise well-designed LDPC code.
In fact, one of the greatest obstacles to the development of a general theory of LDPC codes in the quantum setting is the difficulty of avoiding cycles of length four
(See, for example, \cite{MMM, PC, COT, Hagiwara}).
In order to improve error correction performance, we generally only treat LDPC codes with girth at least six.

The \textit{weight} of a row or column of a binary matrix is its Hamming weight, that is, the number of ones in it.
An LDPC code is \textit{regular} if its parity-check matrix $H$ has constant row and column weights, and \textit{irregular} otherwise.
Regular LDPC codes are known to be able to achieve high error correction performance.
Irregular LDPC codes allow the code designer to optimize characteristics of performance by a careful choice of row weights and column weights \cite{LMSS, RU, RSU}.

We now define several combinatorial structures, which we will need in Subsection \ref{principles} and the subsequent sections.
For additional facts and design theoretical results, the interested reader is referred to \cite{BJL}.

An \textit{incidence structure} is an ordered pair $(V,{\mathcal B})$ such 
that $V$ is
a finite set of \textit{points}, and ${\mathcal B}$ is a family
of subsets of $V$, called \textit{blocks}.
A {\it point-by-block incidence matrix} of an
incidence structure  $(V,{\mathcal B})$ is
a binary $v \times b$ matrix $H = (h_{i,j})$ in which
rows are indexed by points, columns are indexed by blocks,
and $h_{i,j}=1$ if the $i$th point is contained in the $j$th block,
and $h_{i,j}=0$ otherwise.
A {\it block-by-point incidence matrix} of  $(V,{\mathcal B})$ is the transposed point-by-block incidence matrix $H^T$.

Any LDPC code can be associated with an incidence structure by interpreting 
its parity-check matrix as an incidence matrix.
The converse also holds as long as the considered incidence matrix is sparse.

The current paper will focus on incidence structures which have been extensively studied in combinatorics.
This allows us to effectively exploit combinatorial design theory to develop a framework for designing
entanglement-assisted quantum LDPC codes.

A $2$-$(v,\ka,\lambda)$ \textit{design} is an incidence structure $(V, {\mathcal B})$, where $V$ is a set of cardinality $v$ and
${\mathcal B}$ is a family of $\ka$-subsets of $V$
such that each pair of points is contained in exactly $\lambda$ blocks.
We will refer to the parameters $v$, $\ka$, and $\lambda$ as
the \textit{order}, \textit{block size}, and \textit{index} of a $2$-design.
Note that the block size of a $2$-design is usually written as $k$ in the combinatorial literature.
To avoid any confusion with the dimension of a code, we use $\ka$ instead.

The number $b = |{\mathcal B}|$ of blocks in a $2$-$(v,\ka,\lambda)$ design is determined by the design parameters:
\begin{equation}\label{nb}
b = |{\mathcal B}| = \frac{v(v-1)}{\ka(\ka-1)}\lambda.
\end{equation}
A $2$-design is called \textit{symmetric} if $b=v$.

Every point of a $2$-$(v,\ka,\lambda)$ design occurs in exactly $r$ blocks, where
\begin{equation}\label{rn}
r = \frac{v-1}{\ka-1}\lambda.
\end{equation}
The number $r$ is called the {\it replication number} of the design.
A point-by-block incidence matrix $H$ of a 2-$(v,\mu,\lambda)$
design satisfies the equation
\begin{equation}
\label{hht}
HH^T =(r-\lambda)I +\lambda J,
\end{equation}
where $I$ is the identity matrix and $J$ is the $v \times v$ all-one matrix.
Since $r$ and $b$ are integers, it follows that the following two conditions
\begin{eqnarray}\label{nc:bibd}
\left.\begin{aligned}
\lambda(v-1) &\equiv 0 \pmod{\ka -1}, \\
\lambda v(v-1) &\equiv 0 \pmod{\ka(\ka-1)}
\end{aligned}\right.
\end{eqnarray}
are necessary conditions for the existence of a $2$-$(v,\ka,\lambda)$ design.

If the block size $\ka$ and index $\lambda$ are relatively small, an incidence matrix of a $2$-$(v,\ka,\lambda)$ design is sparse.
Hence, a point-by-block incidence matrix of a $2$-$(v,\ka,\lambda)$ design can be viewed as a parity-check matrix $H$ of a regular LDPC code
with constant row weight $r$ and constant column weight $\ka$.
Similarly, a block-by-point incidence matrix defines a code with constant row weight $\ka$ and constant column weight $r$.
In this paper, incidence matrices will generally be point-by-block unless it is specifically noted otherwise.
In the cases when block-by-point matrices are desirable, the notation $H\tr$ will be used.

A substantial part of this paper deals with one of the most fundamental 
incidence structures in combinatorial design theory.
A \textit{Steiner $2$-design}, denoted by $S(2,\ka,v)$, is a $2$-$(v,\ka,1)$ design.
A \textit{Steiner triple system} of order $v$, denoted by STS$(v)$,
is a Steiner 2-design with block size three.
The $S(2,\ka,v)$s are \textit{trivial} Steiner $2$-designs if $v \leq \ka$.
We generally do not consider trivial designs to be Steiner $2$-designs unless they play an important role.

It is easy to see that both point-by-block and block-by-point incidence matrices 
of an $S(2,\ka,v)$ give regular LDPC codes with girth six 
(see, for example, \cite{JohnsonThesis2}).

\subsection{\label{principles}General combinatorial constructions}
In this subsection we present a general framework for designing entanglement-assisted quantum LDPC codes based on combinatorial design theory.
Specialized construction methods for desirable EAQECCs in this framework will be illustrated in Section \ref{finite}.

The following propositions are derived from Theorem \ref{thm:css} by using incidence matrices as parity-check matrices of binary LDPC codes.
\begin{proposition}\label{basicconst1}
Let $H$ be a point-by-block incidence matrix of an incidence structure $(V, {\mathcal B})$.
Then there exists a $[[|{\mathcal B}|,|{\mathcal B}|-2\rk{H}+\rk{HH^T};\rk{HH^T}]]$ \textup{EAQECC}.
\end{proposition}
\begin{proposition}\label{basicconst2}
Let $H^T$ be a block-by-point incidence matrix of an incidence structure $(V, {\mathcal B})$.
Then there exists a $[[|V|,|V|-2\rk{H}+\rk{H^TH};\rk{H^TH}]]$ \textup{EAQECC}.
\end{proposition}

We employ the following two theorems.
\begin{theorem}[Hillebrandt \cite{hillebrandt}]\label{bound:rkSteiner}
The rank of an incidence matrix $H$
of an $S(2,\ka,v)$ satisfies the following  inequalities:
\[ \left\lceil \frac{1}{2}+\sqrt{\frac{1}{4}+\frac{(v-1)(v-\ka)}{\ka}}\right\rceil \leq \rk{H} \leq v.\]
\end{theorem}
\begin{theorem}[Hamada \cite{Ham73}]\label{value:rkSteiner}
If $H$ is an incidence matrix of an $S(2,\ka,v)$ with even replication number 
$r=\frac{v-1}{\ka-1}$ then
\[\rk{H} = \left \{\begin{array}{ccl}
v-1 & \mbox{when} & \ka \mbox{ is even,}\\
v \mbox{ or } v-1 & \mbox{when} & \ka \mbox{ is odd.}
\end{array} \right . \]
\end{theorem}

We now give three simple constructions by applying Propositions \ref{basicconst1} and \ref{basicconst2} to incidence matrices of Steiner $2$-designs.
These constructions will be specialized and modified to give desirable codes.
\begin{theorem}[High-Rate 1-Ebit Code]\label{const1}
Let $H$ be a point-by-block incidence matrix of an $S(2,\ka,v)$. Suppose $r = \frac{v-1}{\ka-1}$ is odd.
Then $H$ has row weight $r$, column weight $\ka$, girth 6, and
the corresponding $[[n,k;c]]$ \textup{EAQECC} satisfies the following conditions:
\[n = \frac{v(v-1)}{\ka(\ka-1)},\]
\[\frac{vr}{\ka}- 2v + 1\leq k \leq  \frac{vr}{\ka}- 2\left\lceil \frac{1}{2}+\sqrt{\frac{1}{4}+\frac{(v-1)(v-\ka)}{\ka}}\right\rceil + 1,\]
\[c=1.\]
\end{theorem}
\Proof
By Proposition \ref{basicconst1} and Theorem \ref{bound:rkSteiner}, it suffices to prove that $\rk{HH^T}=1$.
Because $r$ is odd, Equation (\ref{hht}) reduces to $HH^T = J$, which implies that the rank of $HH^T$ is equal to one.
\qed

\begin{theorem}[High-Rate High-Consumption Code]\label{const2}
Let $H$ be a point-by-block incidence matrix of an $S(2,\ka,v)$. Suppose $r = \frac{v-1}{\ka-1}$ is even.
Then $H$ has row weight $r$, column weight $\ka$, girth 6, and
the corresponding $[[n,k;c]]$ \textup{EAQECC} satisfies the following conditions:
\[n = \frac{v(v-1)}{\ka(\ka-1)},\]
\[k = \left \{\begin{array}{ccl}
\frac{vr}{\ka}-v+1 & \mbox{when} & \ka \mbox{ is even,}\\
\frac{vr}{\ka}-v+1 \mbox{ or } \frac{vr}{\ka}-v-1 & \mbox{when} & \ka \mbox{ is odd,}
\end{array} \right . \]
\[c = v-1.\]
\end{theorem}
\Proof
By Proposition \ref{basicconst1} and Theorem \ref{value:rkSteiner}, it suffices to prove that $\rk{HH^T}=v-1$.
Because $r$ is even, Equation (\ref{hht}) reduces to
\[HH^T = \left[\begin{array}{cccc}
0 & 1 & & 1 \\
1 & 0 & \cdots & 1 \\
& \vdots & \ddots & \vdots\\
1 & 1 & \cdots & 0
\end{array}\right],
\]
that is, a matrix containing zeros on the diagonal and ones in the other entries.
Because $r =  \frac{v-1}{\ka-1}$ is even, $v$ is odd. Hence, we have $\rk{HH^T} = v-1$ as desired.
\qed

\begin{theorem}[Low-Rate High-Redundancy Code]\label{const3}
Let $H^T$ be a block-by-point incidence matrix of an $S(2,\ka,v)$.
Then $H$ has row weight $\ka$, column weight $r$, girth 6, and
the corresponding $[[n,k;c]]$ \textup{EAQECC} satisfies the following conditions:
\[n = v,\]
\[k \leq v - 2\left\lceil \frac{1}{2}+\sqrt{\frac{1}{4}+\frac{(v-1)(v-\ka)}{\ka}}\right\rceil + c,\]
\[c \geq 1.\]
\end{theorem}
\Proof
Let $H^T$ be a block-by-point incidence matrix of an $S(2,\ka,v)$.
Since any non-trivial $S(2,\ka,v)$ contains a pair of blocks that share exactly one point,
we have $\rk{H^TH} \geq 1$.
Applying Proposition \ref{basicconst2} to Theorem \ref{bound:rkSteiner} completes the proof.\qed

It is worth mentioning that a weaker version of Theorem \ref{const1} was used in the context of integrated optics and photonic crystal technology \cite{Djordjevic4}.
Also notable is that Theorems \ref{const1} and \ref{const2} can be easily extended to the case when preexisting entanglement is not available.
For example, quantum LDPC codes that do not require entanglement can be obtained by
applying the extra column method used in Construction U in \cite{MMM} and the CSS construction to $S(2,\mu,v)$s
in the same manner as in Proposition \ref{basicconst1}.
Aly's construction for quantum LDPC codes \cite{Aly} is a special case of this extended method.
Djordjevic's construction for quantum LDPC codes \cite{Djordjevic} can be obtained by applying the CSS construction to $2$-designs of even index
in the same way as in Proposition \ref{basicconst1}.

The existence of $2$-designs is discussed in Appendix \ref{appendix1},
which provides Steiner $2$-designs necessary to obtain several infinite families
of new entanglement-assisted quantum LDPC codes from Theorems \ref{const1}, \ref{const2}, and \ref{const3}.
Before applying our theorems to specific $S(2,\ka,v)$s,
we explore general characteristics of our EAQECCs and further develop methods for designing desirable codes.

Theorem \ref{const1} yields entanglement-assisted quantum LDPC codes
with very high net rates and various lengths while requiring only one ebit.
Theorem \ref{const2} gives codes which have very high net rates and naturally take advantage of larger numbers of ebits
when there is an adequate supply of entanglement.
Since $\rk{HH^T} \leq \rk{H}$ holds for any parity-check matrix $H$,
the required amounts of entanglement of high rate codes in Theorem \ref{const2} are expected to be relatively low
when compared with randomly chosen codes of the same lengths.
Theorem \ref{const3} generates entanglement-assisted quantum LDPC codes which can correct many quantum errors
by taking advantage of the higher redundancy.
The high error correction performance of these codes will be demonstrated in simulations in Section \ref{sim}.

When a parity-check matrix $H$ of an $S(2,\ka,v)$ is of full rank $v$,
the corresponding classical LDPC code in Theorems \ref{const1} and \ref{const2} achieves an upper bound on the rate for an LDPC code with girth six.
\begin{theorem}[MacKay and Davey \cite{MD}]\label{ratebound}
Let $H$ be a $v \times n$ parity-check matrix of a classical regular \textup{LDPC} code of length $n$, column weight $\ka$, and girth $6$. Let also $\rk{H} = v$.
Then it holds that $n \leq \frac{v(v-1)}{\ka(\ka-1)}$, where equality holds if and only if $H$ is an incidence matrix of an $S(2,\ka,v)$.
\end{theorem}
It follows that EAQECCs based on Steiner 2-designs
achieve the highest possible net rates for quantum LDPC codes
with girth at least six constructed from full rank parity-check matrices
with constant column weights
through the CSS construction.

The rank of an incidence matrix of an $S(2,\ka,v)$ may not be full depending on the structure of the design.
If one wishes a parity-check matrix to be regular and full rank at the same time,
it is important to choose an $S(2,\ka,v)$ with a full rank incidence matrix.
This can always be done for the case when $\ka = 3$ except for $v =7$ \cite{DHV}.
For a more detailed treatment of the ranks of $S(2,\ka,v)$s, we refer the reader to \cite{Ham73, Ham68, AK}.

In general, the code minimum distance plays less of a role in the performance
of sum-product decoding than maximum likelihood decoding \cite{MMM}.
Therefore, we explore in detail the distance $d$ of $[[n,k,d;c]]$ EAQECCs based on LDPC codes
only when it is of great theoretical interest.
Because codes derived from finite geometries are of great importance in coding theory,
the distances of EAQECCs obtained from finite geometries will be investigated in detail in Section \ref{finite}.

Here we briefly review the minimum distances of LDPC codes based on Steiner $2$-designs.
A pair of $S(2,\ka,v)$s which are not mutually isomorphic may give different minimum distances.
The tightest known upper and lower bounds on the minimum distance of an LDPC code based on an STS$(v)$
can be found in the very large scale integration (VLSI) literature
as bounds on even-freeness.
\begin{theorem}[Fujiwara and Colbourn \cite{FC}]\label{minimumSTS}
The minimum distance $d$ of a classical binary linear code
whose parity-check matrix forms an incidence matrix of a non-trivial \textup{STS}$(v)$ satisfies $4 \leq d \leq 8$.
\end{theorem}
A carefully chosen triple system can have a good topological structure which gives good decoding performance.
If conditions require larger minimum distances, the code designer may use either block-by-point incidence matrices, or $S(2,\ka,v)$s of larger block sizes.
For known results on minimum distances, girths, and related characteristics of LDPC codes based on combinatorial designs,
the reader is referred to \cite{CF, FC, JohnsonThesis} and references therein.

In what follows, we describe general guidelines for designing entanglement-assisted quantum LDPC codes
with desired parameters and properties by exploiting codes we have presented in this section.

We first consider an $[[n,k;c]]$ EAQECC requiring only a small amount of entanglement.
The extreme case is when $c = 1$. The following theorem gives infinitely many such EAQECCs
having extremely high rates and low decoding complexity.

\begin{theorem}\label{asymptoticconst1}
Let $v$ and $\ka$ be positive integers satisfying $v-1 \equiv 0 \pmod{\ka -1}$ and $v(v-1) \equiv 0 \pmod{\ka(\ka-1)}$.
Suppose also that $\frac{v-1}{\ka-1}$ is odd.
Then for all sufficiently large $v$ and some $k$ satisfying the condition of Theorem \ref{const1}, there exists an $[[\frac{v(v-1)}{\ka(\ka-1)},k;1]]$ \textup{EAQECC}.
\end{theorem}
\Proof
Use Theorem \ref{asymptotic} from Appendix A, which guarantees
the existence of an $S(2,\mu,v)$ for all sufficiently large $v$,
and apply Theorem \ref{const1}.
\qed

Similarly, applying Theorem \ref{const1} to known $S(2,\ka,v)$s with small $v$ discussed in Appendix \ref{appendix1} gives $[[n,k;1]]$ EAQECCs of shorter length $n$.

In general, the error floor of a well-designed LDPC code is not dominated by low-weight codewords.
Nonetheless, it is desirable to carefully choose an $S(2,\ka,v)$ when applying our simple constructions
so that the resulting code has a promising topological structure.
While incidence matrices of $S(2,\ka,v)$s have long been investigated in various fields,
it appears to be difficult to achieve the known upper bounds on the minimum distance of an LDPC code based on an incidence matrix of an $S(2,\ka,v)$.
In fact, it is conjectured that the known upper bounds are generally not achievable even for the case $\ka = 3$ \cite{CF}.

An STS is $4$-\textit{even-free} (or \textit{anti-Pasch}) if its incidence matrix gives a classical LDPC code with minimum distance five or greater.
A $4$-even-free STS$(v)$ exists for all $v \not = 7, 13$ satisfying the necessary conditions (\ref{nc:bibd}) \cite{GGW}.
It is conjectured that an incidence matrix of a $4$-even-free STS$(v)$ gives the largest possible minimum distance \cite{CF}.
\begin{theorem}\label{stseaqecc}
There exists a $[[\frac{v(v-1)}{6},k,d;1]]$ \textup{EAQECC} with $k \geq \frac{v(v-1)}{6}-2v+1$ and $d \geq 5$ for every $v \equiv 3, 7 \pmod{12}$ except for $v=7$.
\end{theorem}
\Proof
If  $v \equiv 3, 7 \pmod{12}$, then the replication number of an STS$(v)$ is odd. Applying Theorem \ref{const1} to a $4$-even-free STS$(v)$ completes the proof.\qed

A block-by-point incidence matrix of a symmetric $S(2,\ka,v)$ can also be viewed as a point-by-block incidence matrix of a Steiner $2$-design of the same parameters \cite{HandbookCD}.
Hence, Theorems \ref{const1} and \ref{const3} can overlap when symmetric designs are employed.
This special case gives the EAQECCs with $c=1$ and good error correction performance originally presented in \cite{HYH}.
For completeness, we give a simple proof by using the following two theorems.
\begin{theorem}\label{th:existenceplane}
For every integer $t \geq 1$ there exists a symmetric $S(2,2^t+1,4^t+2^t+1)$ whose incidence matrix $H$ satisfies $\rk{H} = 3^t +1$. 
\end{theorem}
\Proof Take as $S(2,2^t+1,4^t+2^t+1)$ the Desarguesian projective plane of 
order $2^t$, whose incidence matrix has rank $3^t +1$ \cite{GM}.\qed

\begin{theorem}[Calkin, Key, and de Resmini \cite{CKR}]\label{th:distanceplane}
Let $H^T$ be a block-by-point incidence matrix of 
a symmetric $S(2,2^t+1,4^t+2^t+1)$ being the Desarguesian projective plane
$PG(2,2^t)$.
Then $H^T$ defines a classical binary linear $[4^t+2^t+1, 4^t+2^t-3^t,2^t+2]$ code.
\end{theorem}

Now as a corollary of Theorems \ref{const1} and \ref{const3} and the preceding two theorems, 
we obtain the following result.
\begin{theorem}\label{plane}
For every integer $t \geq 1$ there exists a $[[4^t+2^t+1, 4^t+2^t-2\cdot3^t,2^t+2;1]]$ \textup{EAQECC}.
\end{theorem}
EAQECCs of this kind can be seen as quantum analogues of special Type I PG-LDPC codes,
which have notable error correction performance in the classical setting \cite{KLF, TXKLA, TXLA}.
Because of the direct correspondence between entanglement-assisted quantum codes and classical codes,
these EAQECCs inherit excellent error correction performance while consuming only one initial ebit.
We will further investigate entanglement-assisted quantum LDPC codes based on $S(2,\ka,v)$s
with large minimum distances in Section \ref{finite}.

Next we present general combinatorial methods for designing EAQECCs with relatively small $c$ and better error correction performance.
The main idea is that we discard some columns from an incidence matrix of an $S(2,\ka,v)$
and then apply Proposition \ref{basicconst1} as we did in Theorem \ref{const1}.
Our methods encompass the rate control technique for classical LDPC codes proposed in \cite{JW2} as a special case.

Let $(V,{\mathcal B})$ be an $S(2,\ka,v)$. Take two subsets $V' \subsetneq V$ and ${\mathcal B}' \subsetneq {\mathcal B}$.
The pair $(V', {\mathcal B}')$ is called a \textit{proper subdesign} of block size $\ka$ if it is an $S(2,\ka,|V'|)$.
Since we do not consider other kinds of subdesigns, we simply call a proper subdesign $(V', {\mathcal B}')$ of block size $\ka$ a subdesign.
A pair of subdesigns $(V',{\mathcal B}')$ and $(V'',{\mathcal B}'')$ of an $S(2,\ka,v)$ are \textit{point-wise disjoint} if $V' \cap V'' = \emptyset$.

\begin{theorem}\label{ratecontrol}
Let $(V, {\mathcal B})$ be an $S(2,\ka,v)$ with odd $r = \frac{v-1}{\ka-1}$.
Assume that $(V, {\mathcal B})$ contains $j$ point-wise mutually disjoint subdesigns $(V_i, {\mathcal B}_i)$,
$1 \leq i \leq j$, such that $\bigcup_{i=1}^jV_i \subsetneq V$ and each $(V_i, {\mathcal B}_i)$ has odd replication number.
Then there exists an $[[n,k;c]]$ \textup{EAQECC} satisfying the following conditions:
\[n = \frac{v(v-1)}{\ka(\ka-1)}-|\bigcup {\mathcal B}_i|,\]
\[c=j+1.\]
\end{theorem}
\Proof
Take an arbitrary incidence matrix $H$ of an $S(2,\ka,v)$ with odd $r$.
Delete $j$ point-wise mutually disjoint subdesigns $(V_i, {\mathcal B}_i)$ each of which has odd replication number.
It is always possible to reorder the rows and columns of the resulting incidence matrix $H'$ such that $H'H'^T$ has the form:
\[H'H'^T = \left[\begin{array}{cccc}
J & J & & J \\
J & 0_1 & \cdots & J \\
& \vdots & \ddots & \vdots\\
J & J & \cdots & 0_j
\end{array}\right]
\]
where $0_i$ is a $|V_i|\times|V_i|$ zero matrix and each $J$ is an all-one matrix of appropriate size.
It is easy to see that $\rk{H'H'^T} = j+1$.
Applying Proposition \ref{basicconst1} to $H'$ completes the proof.
\qed

Deleting subdesigns always shortens the length of the corresponding code. Discarding columns will not decrease the minimum distance or the girth.
The rank of the parity-check matrix is unlikely to change.
In this sense, we expect EAQECCs obtained through subdesign deletion to have better error correction performance than the original code.
We will demonstrate this effect in simulations in Section \ref{sim}.

In general, deleting a subdesign makes a parity-check matrix slightly irregular.
If this irregularity is not desirable because of particular circumstances or conditions,
it can be alleviated by discarding more point-wise disjoint subdesigns.
In fact, if we delete subdesigns of the same order such that each point belongs to one deleted subdesign, we obtain a regular parity-check matrix again.
The following construction demonstrates this.

Let $(V, {\mathcal B})$ be an $S(2,\ka,v)$ and
${\mathcal S}$ a set of Steiner $2$-designs $(V_i,{\mathcal B_i}), 1 \leq i \leq |{\mathcal S}|$, where
$V_1,\dots,V_{|{\mathcal S}|}$ partition $V$, that is, $\bigcup V_i = V$ and $V_i \cap V_j = \emptyset$ for all $i \not= j$.
Then ${\mathcal S}$ is called a \textit{Steiner} \textit{spread} in $(V, {\mathcal B})$
if each $(V_i,{\mathcal B_i})$ forms a subdesign $S(2,\ka,|V_i|)$ of $(V, {\mathcal B})$.

\begin{theorem}\label{ratecontrolspread}
Let $(V, {\mathcal B})$ be an $S(2,\ka,v)$ with odd replication number $r = \frac{v-1}{\ka-1}$.
Assume that $(V, {\mathcal B})$ contains a Steiner spread ${\mathcal S}$,
where each subdesign $(V_i, {\mathcal B}_i)$ has odd replication number.
Then there exists an $[[n,k;c]]$ \textup{EAQECC} satisfying the following conditions:
\[n = \frac{v(v-1)}{\ka(\ka-1)}-|\bigcup {\mathcal B}_i|,\]
\[
c = \left \{\begin{array}{ccl}
|{\mathcal S}|-1 & \mbox{when} & |{\mathcal S}| \mbox{ is odd,}\\
|{\mathcal S}| & \mbox{when} & |{\mathcal S}| \mbox{ is even.}
\end{array} \right .\]
Moreover, if $|V_i| = |V_{i'}| = w$ for all $i$ and $i'$,
then the parity-check matrix of the corresponding \textup{LDPC} code is regular
and has row weight $r - \frac{w-1}{\ka-1}$ and column weight $\ka$.
\end{theorem}
\Proof
Let $H$ be an incidence matrix of an $S(2,\ka,v)$ with odd $r$ which contains a Steiner spread ${\mathcal S}$.
Delete all members of the Steiner spread from $(V, {\mathcal B})$.
By following the same argument as in the proof of Theorem \ref {ratecontrol},
it is straightforward to see that $\rk{HH^T} = |{\mathcal S}| -1$ when $|{\mathcal S}|$ is odd, and $|{\mathcal S}|$ otherwise.
If $|V_i| = |V_{i'}| = w$ for all $i$ and $i'$, each subdesign has the same replication number $ \frac{w-1}{\ka-1}$.
Hence, the resulting code is regular.
\qed

When there is an adequate supply of entanglement,
it may be acceptable to exploit a relatively large amount of entanglement
to improve error correction performance while keeping similar characteristics of high rate codes.
Deleting an $S(2,\ka,w)$ with even replication number $\frac{w-1}{k-1}$ increases the required amount of entanglement to a slightly larger extent.

\begin{theorem}\label{ratecontroloddr}
Let $(V, {\mathcal B})$ be an $S(2,\ka,v)$ with odd replication number $r = \frac{v-1}{\ka-1}$.
Assume that $(V, {\mathcal B})$ contains $j$ point-wise mutually disjoint subdesigns $(V_i, {\mathcal B}_i)$,
$1 \leq i \leq j$, such that $\bigcup_{i=1}^jV_i \subseteq V$ and each $(V_i, {\mathcal B}_i)$ has even replication number.
Then there exists an $[[n,k;c]]$ \textup{EAQECC} satisfying the following conditions:
\[n = \frac{v(v-1)}{\ka(\ka-1)}-|\bigcup {\mathcal B}_i|,\]
\[c=\sum_{i=1}^{j}{(|V_i|-1)}+1.\]
Moreover, if the subdesigns $(V_i, {\mathcal B}_i)$ for $1 \leq i \leq j$ form a Steiner spread
with $|V_i| = |V_{i'}| = w$ for all $i$ and $i'$, then the parity-check matrix of the corresponding \textup{LDPC} code is regular
and has row weight $r - \frac{w-1}{\ka-1}$ and column weight $\ka$.
\end{theorem}
\Proof
Take an arbitrary incidence matrix $H$ of an $S(2,\ka,v)$ with odd $r$.
Delete $j$ point-wise mutually disjoint subdesigns $(V_i, {\mathcal B}_i)$ each of which has even replication number.
If $\bigcup_{i=1}^jV_i \subsetneq V$, it is always possible to reorder the columns of the resulting incidence matrix $H'$ such that $H'H'^T$ is of the form:
\[H'H'^T = \left[\begin{array}{cccc}
J & J & & J \\
J & I_1 & \cdots & J \\
& \vdots & \ddots & \vdots\\
J & J & \cdots & I_j
\end{array}\right]
\]
where $I_i$ is the $|V_i| \times |V_i|$ identity matrix and each $J$ is an all-one matrix of appropriate size.
Because each $I_i$ has $V_i$ independent rows and each $|V_i|$ is odd, $\rk{H'H'^T} = \sum_{i=1}^{j}{(|V_i|-1)}+1$.
Applying Proposition \ref{basicconst1} to $H'$ gives $c=\sum_{i=1}^{j}{(|V_i|-1)}+1$. If $\bigcup_{i=1}^jV_i = V$,
we have identity matrices across the diagonal of $H'H'^T$.
Hence, we have $c=\sum_{i=1}^{j}{(|V_i|-1)}+1$ again.
If each $V_i$ is of the same size, it is straightforward to see that the resulting code is regular.\qed

When irregularity in a parity-check matrix is acceptable or favorable,
the code designer can combine the techniques of Theorems \ref{ratecontrol}, \ref{ratecontrolspread}, and \ref{ratecontroloddr}.
The required amount of entanglement is readily computed by the same argument as above.

In general, subdesign deletion changes the parameters of a code in a gradual manner.
Hence, these techniques are also useful when one would like an EAQECC of specific length or dimension.
While we only employed Theorem \ref{const1} in the above arguments,
Theorem \ref{const2} can also be used in a straightforward manner
to fine-tune the parameters of EAQECCs.

In order to exploit the subdesign deletion techniques, one needs Steiner $2$-designs having subdesigns or
preferably Steiner spreads of appropriate sizes.
We conclude this section with a brief review of known general results and useful theorems for finding $S(2,\ka,v)$ with subdesigns and Steiner spreads.
For a more thorough treatment, the reader is referred to \cite{HandbookCD, BJL} and references therein.

The well-known Doyen-Wilson theorem \cite{DW} states that one can always find an STS$(v)$ containing an STS$(w)$ as a subdesign
as long as both $v$ and $w$ satisfy the necessary conditions for the existence of an STS and $v \geq 2w+1$.
The following is a general asymptotic theorem on Steiner $2$-designs having subdesigns.
\begin{theorem}[Fujiwara \cite{halving}]\label{subasymptotic}
Let $\ka \geq 2$ be a positive integer and $w \equiv 1$ {\rm(mod $\ka(\ka-1)$)}. 
Then there exist a constant number $w_0$ depending on $\ka$, 
and a constant number $v_0$ depending on $w$ and $\ka$ 
such that if $w > w_0$ and $v>v_0$ satisfies the conditions 
$v-1 \equiv 0$ {\rm(mod $\ka-1$)} and $v(v-1) \equiv 0$ {\rm(mod $\ka(\ka-1)$)}, 
then there exists an $S(2,\ka,v)$ having an $S(2,\ka,w)$ as a subdesign. 
\end{theorem}
Theorem \ref{subasymptotic} states that one can always find an $S(2,\ka,v)$ having an $S(2,\ka,w)$ as a subdesign
as long as $v$ is a sufficiently large integer satisfying the necessary conditions (\ref{nc:bibd})
and $w$ is a sufficiently large integer satisfying $w \equiv 1$ {\rm(mod $\ka(\ka-1)$)}.

Steiner spreads are closely related to a special kind of combinatorial design.
A {\it group divisible design} (GDD) with {\it index} one is a triple 
$(V, {\mathcal G}, {\mathcal B})$, where
\begin{enumerate}
	\item[(i)] $V$ is a finite set of elements called \textit{points}, 
        \item[(ii)] ${\mathcal G}$ is a family of subsets of $V$, called {\it groups}, 
                which partition $V$, 
        \item[(iii)] ${\mathcal B}$ is a collection of subsets of $V$, called 
                {\it blocks}, such that every pair of points from distinct 
                groups occurs in exactly one block, 
        \item[(iv)] $|G \cap B| \leq 1$ for all $G \in {\mathcal G}$ and $B 
                \in {\mathcal B}$.
\end{enumerate}
If all groups are of the same size $g$, all blocks are of the same size $\ka$, and 
$|{\mathcal G}| = t$, one refers to the design as a 
$\ka$-GDD of {\it type} $g^t$.

\begin{theorem}\label{GDDspread}
The existence of an $S(2,\ka,g)$ and a $\ka$-\textup{GDD} 
$(V, {\mathcal G}, {\mathcal B})$ of {\it type} $g^t$ with index one
implies the existence of an $S(2,{\ka},gt)$ having a Steiner 
spread ${\mathcal S}$, where each member of ${\mathcal S}$ is an $S(2,\ka,g)$.
\end{theorem}
\Proof
Let $(V, {\mathcal G}, {\mathcal B})$ be a $\ka$-\textup{GDD} of {\it type} $g^t$ with index one and $(V', {\mathcal B}')$ an $S(2,\ka,g)$.
For each $G \in {\mathcal G}$, we construct an $S(2,\ka,g)$, $(G, {\mathcal B}'_G)$,
by mapping each point of $(V', {\mathcal B}')$ to an element of $G$ by an arbitrary bijection $\pi_G : V' \rightarrow G$.
Define ${\mathcal C} = \bigcup_{G \in {\mathcal G}}{\mathcal B}'_G$.
It is straightforward to check that $(V, {\mathcal B}\cup{\mathcal C})$ is an $S(2,\ka,gt)$ having a Steiner spread whose members are all $S(2,\ka,g)$s.
\qed

The above theorem is useful to obtain regular LDPC codes through Theorems \ref{ratecontrolspread} and \ref{ratecontroloddr}
and similar subdesgin deletion techniques based on Theorem \ref{const2}.
One can also modify Theorem \ref{GDDspread} for the case when a GDD has different group sizes by a similar argument.
The existence of GDDs and their constructions have been extensively investigated in combinatorial design theory.
For a comprehensive list of known existence results on GDDs, we refer the reader to \cite{HandbookCD}.

\section{\label{finite}Finite geometry codes}
In this section, we demonstrate applications of our general designing methods
by using combinatorial designs arising from finite geometries.

The classical LDPC codes obtained from finite geometries are known to have remarkable error correction abilities.
By using these codes, we generate infinitely many new high performance entanglement-assisted quantum LDPC codes having numerous Steiner spreads of various sizes.
The various Steiner spreads in each code allow the code designer to flexibly fine-tune the parameters and error correction performance.

This section is divided into three subsections.
Subsection \ref{sub:pg} studies entanglement-assisted quantum LDPC codes of girth six obtained from projective geometries.
Codes based on affine geometries are investigated in Subsection \ref{sub:ag}.
In Subsection \ref{sub:eg} we investigate slightly modified affine geometry codes, called Euclidean geometry codes.
Classical LDPC codes based on these three kinds of finite geometries are called \textit{finite geometry LDPC codes}
or simply \textit{FG-LDPC codes}.

Many of the results presented in this section can also be seen as new results on classical finite geometry LDPC codes.
In particular, properties of finite geometries have been independently studied in the combinatorial literature,
and hence many of the ``known" results are new results in the field of LDPC codes.
For the convenience of the reader,
we summarize our results on fundamental parameters of LDPC codes from finite geometries
in Tables \ref{fig:quantumsummary} and \ref{fig:classicalsummary} in Appendix \ref{appendix2}.
Lengths, dimensions, and minimum distances of the FG-LDPC codes with girth six from
projective geometry $PG(m,q)$, affine geometry $AG(m,q)$, and Euclidean geometry $EG(2,2^t)$ are all determined.
Specifically for EAQECCs based on FG-LDPC codes, we also determine the required amounts of entanglement for most cases.
For a few cases, we give upper bounds on the required amount of entanglement.

\subsection{Projective geometry codes\label{sub:pg}}
We begin with EAQECCs obtained from finite projective geometries.
The use of projective geometries for constructing EAQECCs first appeared
in the work of Hsieh, Yen, and Hsu \cite{HYH}.
This subsection illustrates how our combinatorial framework generalizes 
their method and determines fundamental parameters
of quantum and classical LDPC codes obtained from $PG(m,q)$.

Points of the $m$-dimensional projective geometry $PG(m,q)$ over $\fq$ are the 1-dimensional subspaces of $\fq^{m+1}$.
The $i$-dimensional projective subspaces of $PG(m,q)$ are the $(i+1)$-dimensional vector subspaces of $\fq^{m+1}$.
The points and lines of $PG(m,q)$ form an $S(2,q+1,\frac{q^{m+1}-1}{q-1})$,
denoted by $PG_{1}(m,q)$, having $\frac{(q^{m+1}-1)(q^{m}-1)}{(q^{2}-1)(q-1)}$
blocks and replication number
$\frac{q^{m}-1}{q-1} = q^{m-1} + q^{m-2} + \cdots + q + 1.$

One can obtain two types of EAQECCs from projective geometry designs:
Type II (using a point-by-block incidence matrix) and
Type I (using a block-by-point incidence matrix of the design). 
Applying Proposition \ref{basicconst1} to an incidence matrix of $PG_{1}(m,q)$, we obtain a Type II EAQECC.
This type of EAQECC belongs to the high rate entanglement-assisted quantum LDPC codes given in Theorems \ref{const1} and \ref{const2}.
If we apply Proposition \ref{basicconst2} to a block-by-point incidence matrix, we obtain a Type I EAQECC.
This kind of EAQECC belongs to the high redundancy entanglement-assisted quantum LDPC codes given in Theorem \ref{const3}.

The rank of an incidence matrix determines the dimension of the corresponding FG-LDPC code,
hence it is one of the key values in the quantum setting as well.
Exact values for many sporadic examples have been computed in the fields of quantum and classical LDPC codes.
The following two theorems give the exact rank for all projective geometry designs.
\begin{theorem}[Hamada \cite{Ham68}]\label{thm:hamada}
The rank of $PG_{1}(m,2^{t})$ is given by
%\begin{widetext}
	\begin{eqnarray*}&&\rk{PG_1(m,2^t)} = \varphi(m,2^t) =\\
	&&\sum_{(s_{0}, s_{1}, \ldots, s_{t})}\prod_{j=0}^{t-1}\sum_{i=0}^{L(s_{j+1},s_{j})}l^i{m+1 \choose i}{m+2s_{j+1}-s_{j}-2i \choose m}
	\end{eqnarray*}
%\end{widetext}
	where $l = -1$, the sum is taken over all ordered sets
	$(s_{0}, s_{1}, \ldots, s_{t})$ with $s_{0} = s_{t}$, $s_{j} \in \mathbb{Z}$ such that $0 \leq s_{j} \leq m-1$
	and $0 \leq 2s_{j+1} - s_{j} \leq m+1$ for each $j = 0, \ldots, t-1$, and \[L(s_{j+1},s_{j}) = \left[\frac{2s_{j+1} - s_{j}}{2}\right].\]
\end{theorem}
We will use the notation $\varphi(m,2^t)$ for the rank of $PG_{1}(m,q)$ when $q$ is even, that is, $q = 2^{t}$.
When $q$ is odd, the rank of $PG_{1}(m,q)$
is given by a formula of Frumkin and Yakir \cite{FY}.
\begin{theorem}[Frumkin and Yakir \cite{FY}]\label{pgodddimension}
Let $q$ be odd and $H$ an incidence matrix of the design $PG_{1}(m,q)$ with $v= \frac{q^{m+1}-1}{q-1}$ points. Then $\rk{H} = v-1 = \frac{q^{m+1}-q}{q-1}$.
\end{theorem}
Hence the exact dimensions of the corresponding FG-LDPC codes obtained from projective geometries can be calculated for all cases.

The rank of $PG_{1}(m,2^{t})$ was conjectured by Hamada \cite{Ham73}
to be the lowest rank among all Steiner $2$-designs of the same order and block size.
This has been confirmed in a number of cases, although in general the conjecture is still open.
Thus we expect that the designs $PG_{1}(m,2^t)$ should provide codes with the best possible dimensions
among all non-isomorphic $S(2,2^t+1,\frac{2^{t(m+1)}-1}{2^t-1})$s.

We will now examine the codes obtained from $PG_{1}(m,q)$ in detail. 
This subsection is divided into two portions based on the orientation of the 
incidence matrix.

\subsubsection{Point-by-block (Type II) Projective geometry codes}
In this portion, we consider the EAQECCs corresponding to a point-by-block incidence matrix of $PG_{1}(m,q)$.

We first consider the case $q = 2^t$ for some positive integer $t$.
The following theorem gives an infinite family of entanglement-assisted quantum LDPC codes which consume only one initial ebit and have extremely large net rate.
\begin{theorem}\label{th:pgm2t}
For every pair of integers $t \geq 1$ and $m \geq 2$ there exists an entanglement-assisted quantum \textup{LDPC} codes with girth six whose parameters $[[n,k,d;c]]$ are
\[n = \frac{(2^{t(m+1)}-1)(2^{tm}-1)}{(2^{2t}-1)(2^t-1)},\]
\[k = \frac{(2^{t(m+1)}-1)(2^{tm}-1)}{(2^{2t}-1)(2^t-1)} - 2\varphi(m,2^t) + 1,\]
\[d = 2^t+2, \ \mbox{and}\]
\[c = 1.\]
\end{theorem}

To prove Theorem \ref{th:pgm2t}, we first prove a new result on the distance of EAQECCs obtained from an incidence matrix of $PG_{1}(m,2^{t})$.
We use a special set of lines.
A \emph{dual hyperoval} ${\mathcal H}$ is a set of $q+2$ lines of $PG_1(2,q)$,
such that each point of $PG_1(2,q)$ lies on either zero or two lines of ${\mathcal H}$.
Dual hyperovals exist if and only if $q$ is even.
An example is the set of projective lines with equations
\[\{X_0+\beta X_1+\beta^2 X_2=0:\beta\in\fq\}\cup \{X_1=0\}\cup \{X_2=0\}.\]

\begin{theorem}\label{lem:pgevendist}
Let $H$ be an incidence matrix of $PG_1(m,2^t)$.
The minimum distance of the classical binary linear code with parity-check
matrix $H$ is $2^t+2$.
\end{theorem}
\Proof First, we note that coordinates of the codewords correspond to lines of
the geometry, and a codeword corresponds to a set $S$ of lines in $PG_1(m,2^t)$
such that every point is contained in an even number of lines of $S$.
Assume that $c$ is a non-zero codeword,
and let $\supp(c)$ denote the support of $c$, that is, the set of
indices of the nonzero coordinates of $c$.
Since $c\neq 0$, the support of $c$ contains at least one line $\ell$.
Through each point of $PG(m,2^t)$, there pass an even number of lines
from $\supp(c)$. In particular, each of the $2^t+1$ points on $\ell$ lies on at
least one other line of $\supp(c)$, and all these lines are different as they
have different intersections with $\ell$. Hence, there are at least
$1+(2^t+1)$ lines in $\supp(c)$, that is, minimum distance $d$ is at least $2^t+2$.
Let $\pi$ be a plane in $PG(m,2^t)$ and $S$ the set of the $2^t+2$ lines of a dual hyperoval in $\pi$.
Then $S$ corresponds to a codeword of weight $2^t+2$, hence $d=2^t+2$.\qed

\noindent\textbf{Proof of Theorem \ref{th:pgm2t}.}
Let $H$ be an incidence matrix of $PG_{1}(m,2^{t})$.
The rank of $H$ is $\varphi(m,2^t)$ given by Theorem \ref{thm:hamada}.
The index of $PG_1(m,2^t)$ is one. The replication number is odd.
By Equation (\ref{hht}) and Theorem \ref{const1}, we have $\rk{HH^T} = 1$.
By Theorem \ref{lem:pgevendist}, the minimum distance of the binary linear
code with parity-check matrix $H$ is $2^t+2$.
\qed

Next, we examine EAQECCs obtained from an incidence matrix of $PG_{1}(m,q)$ with $q$ odd.
This case also gives very high rate entanglement-assisted quantum LDPC codes.

\begin{lemma}\label{lem:planecase}
Let $H$ be an incidence matrix of $PG_1(2,q)$, $q$ odd.
Then the classical binary linear code defined by parity-check matrix $H$
consists of only the zero vector and the all-one vector.
\end{lemma}

\Proof This follows directly from Theorem \ref{pgodddimension}.

A \emph{hyperbolic quadric} $Q$ is a substructure
$(\mathcal P,\mathcal L)$ of $PG_1(3,q)$ with $(q+1)^2$ points
and $2(q+1)$ lines, such that each point of $\mathcal P$ lies on
exactly two lines of $\mathcal L$ and every plane
of $PG(3,q)$ contains zero or two lines of $\mathcal L$.
Hyperbolic quadrics exist for every odd prime power $q$.

\begin{theorem}\label{pgdistance2}
Let $H$ be an incidence matrix of $PG_1(m,q)$, $m\ge 3$, $q$ odd.
Then the minimum distance of the classical binary linear code with a
parity-check matrix $H$ is $2(q+1)$.
\end{theorem}

\Proof Let $\Pi$ be a $3$-dimensional subspace of $PG(m,q)$ and
$(\mathcal P,\mathcal L)$ a hyperbolic quadric in $\Pi$.
The set of lines $\mathcal L$  determines a codeword
of weight $2q+2$, since each point of $PG(m,q)$ is contained in zero or
two lines of $\mathcal L$.
Hence minimum distance $d$ is at least $2q+2$.

We show that there are no codewords of weight smaller than $2q+2$.
Assume that there exists a codeword $c$ of weight smaller than $2q+2$,
that is, $\supp(c)$ is a set of less than $2q+2$ lines of $PG(m,q)$, such that
each point lies on an even number of lines of $\supp(c)$. We will show that
for any $2$-dimensional subspace $\pi$ one has
either $|\supp(c)\cap\pi|\le 1$ or $|\supp(c)\cap\pi|\ge q+2$.

First, let $S =\supp(c)\cap\pi=\{\ell_1,\ldots,\ell_i\}$.
For each $j\in\{1,\ldots,i\}$, each of the points on $\ell_j$ has
to lie on at least one other line of $\supp(c)$, and at most $i-1$ of
them can lie on a line of $S$. Hence, at least $q+1-(i-1)$ of them are
lines in $\supp(c)\setminus S$ and since they all have different
intersections with $\pi$, this yields $i(q-i+2)$ lines in $\supp(c)\setminus S$.
Together with the $i$ lines of $S$, we have $$i(q-i+2)+i<2q+2$$ and solving
this quadratic inequality for $i$ gives us that either $i>q+1$ or $i<2$.
Since $i$ is an integer, hence $i\ge q+2$ or $i\le 1$.

Now, let $\ell$ be any line of $\supp(c)$. Each point of $\ell$ must lie on
at least one other line, hence there certainly exist planes $\pi$ with $i\ge 2$,
and we have $i\ge q+2$. Let $\pi$ be such a plane. We will now show that all
lines of $\supp(c)$ are contained in $\pi$. Assume the contrary, that
there exists a line $\ell'\in \supp(c)\setminus S$. Through each of the points
on $\ell'\setminus\pi$, we need at least one other line of $\supp(c)$ which
is not contained in $\pi$. Since there are at least $q$ points on $\ell'\setminus\pi$,
one has
$$|\supp(c)|=|S|+|\supp(c)\setminus S|\ge (q+2)+(1+q)>2q+2,$$
a contradiction.
Hence, $\ell'$ does not exist and $\supp(c)$ is contained within a single plane $\pi$.
However, $\pi$ is a $PG_1(2,q)$ and by Lemma \ref{lem:planecase} we need
 $q^2+q+1>2q+2$ lines in this case, a contradiction.
Hence, there are no codewords of weight less than $2q+2$.\qed

We now give another infinite family of Type II entanglement-assisted quantum LDPC codes.
\begin{theorem}\label{th:pgoddq}
Let $q$ be an odd prime power.
Then for every integer $m \geq 3$ there exists an
entanglement-assisted quantum \textup{LDPC} code with girth six
whose parameters $[[n,k,d;c]]$ are
\[n = \frac{(q^{m+1}-1)(q^{m}-1)}{(q^{2}-1)(q-1)},\]
\[k = \frac{(q^{m+1}-1)(q^{m}-1)}{(q^{2}-1)(q-1)} - 2\frac{q^{m+1}-q}{q-1}+c,\]
\[d = 2q+2,  \ \mbox{and}\]
\[c =  \left \{\begin{array}{ccl}
1 & \mbox{when} & m \mbox{ is odd,}\\
\frac{q^{m+1}-q}{q-1} & \mbox{when} & m \mbox{ is even.}
\end{array} \right.\]
\end{theorem}
\Proof
This follows directly from Proposition \ref{basicconst1} and Theorems \ref{const1}, \ref{pgodddimension}, and \ref{pgdistance2}.
\qed

Therefore in the case where $m$ is odd, we have another infinite class of EAQECCs which consume only one ebit.
If $m$ is even, we obtain infinitely many high rate codes which consume reasonable numbers of ebit.
Tables \ref{fig:pg2paramseven} and \ref{fig:pg2paramsodd} give a sample of the parameters of the Type II codes obtained from $PG_{1}(m,q)$ with $q$ even and $q$ odd respectively.
\begin{table}[h!t]\caption{Sample parameters of Type II $\qecci{n}{k}{d}{c}$ EAQECCs obtained from $PG_{1}(m,q)$, $q$ even.\label{fig:pg2paramseven}}
\begin{ruledtabular}
\begin{tabular}{llllll}
$m$ & $q$ & $n$ & $k$ & $d$ & $c$ \\\hline
3 & 2 & 35 & 14 & 4 & 1\\
4 & 2 & 155 & 104 & 4 & 1\\
5 & 2 & 651 & 538 & 4 & 1\\
6 & 2 & 2667 & 2428 & 4 & 1\\
3 & 4 & 357 & 236 & 6 & 1\\
4 & 4 & 5795 & 5204 & 6 & 1\\
2 & 8 & 73 & 18 & 10 & 1\\
3 & 8 & 4745 & 3944 & 10 & 1
\end{tabular}
\end{ruledtabular}
\end{table}
\begin{table}[h!t]\caption{Sample parameters of Type II $\qecci{n}{k}{d}{c}$ EAQECCs obtained from $PG_{1}(m,q)$, $q$ odd.\label{fig:pg2paramsodd}}
\begin{ruledtabular}
\begin{tabular}{llllll}
$m$ & $q$ & $n$ & $k$ & $d$ & $c$ \\\hline
3 & 3 & 130 & 53 & 8 & 1 \\
3 & 5 & 806 & 497 & 12 & 1\\
3 & 7 & 2850 & 2053 & 16 & 1\\
4 & 3 & 1210 & 1090 & 8 & 120
\end{tabular}
\end{ruledtabular}
\end{table}

In the reminder of this portion, we examine Steiner spreads of projective geometry designs.
These substructures can be used in Theorems \ref{ratecontrol}, \ref{ratecontrolspread}, and \ref{ratecontroloddr}
and their analogous techniques based on Theorem \ref{const2} to fine-turn the rates and distances of the EAQECCs.

An \emph{$s$-spread} of $PG(m,q)$ is a set of $s$-dimensional projective subspaces which partition the points of the geometry.
In other words, an $s$-spread consists of a set of $(s+1)$-dimensional vector subspaces of $\fq^{m+1}$ which contain every nonzero vector exactly once.
It is known that $PG(m,q)$ admits an $s$-spread if and only if $s+1$ divides $m+1$ (see \cite{Segre} and \cite[p. 29]{Dembowski}).

Take $PG_{1}(m,q)$ and suppose $s \geq 2$ is chosen so that $s+1$ divides $m+1$. Then an $s$-spread of $PG(m,q)$ exists.
Each $s$-dimensional subspace in the spread contains an isomorphic copy of $PG_{1}(s,q)$,
and hence this forms a Steiner spread.
Note that the blocks of $PG_{1}(s,q)$ have size $q+1$ and are also blocks of $PG_{1}(m,q)$. Therefore we have the following result.
\begin{theorem}
Let $s$, $m \geq 1$ be positive integers such that $s+1$ divides $m+1$.
Then $PG_1(m,q)$ contains $\frac{q^{m+1}-1}{q^{s+1}-1}$ disjoint copies of $PG_{1}(s,q)$ whose point sets partition the point of $PG_1(m,q)$.
\end{theorem}
Thus, we can find a set of disjoint subdesigns which partition the points of $PG_{1}(m,q)$ whenever $m+1$ has a nontrivial factor.
Naturally, we may further sub-divide each subdesign of dimension $s$ into smaller subdesigns, based on the nontrivial factors of $s+1$.
Hence, the $S(2,\ka,v)$s from $PG_{1}(m,q)$ are very flexible in that they have Steiner spreads of various sizes.

In general, the length, dimension, required ebits, and rate each change gradually as we delete subdesigns in a Steiner spread.
The minimum distance and rank are either remain the same or improve slightly.
Table \ref{fig:pg152subdesigns} lists the example parameters of EAQECCs created by deleting subdesigns from $PG_{1}(5,2)$.
The first and last rows correspond to regular LDPC codes.
\begin{table}[h!t]
\caption{Summary of Type II codes obtained by deleting a Steiner spread of subdesigns isomorphic to $PG_{1}(2,2)$ from $PG_{1}(5,2)$.\label{fig:pg152subdesigns}}
\begin{ruledtabular}
\begin{tabular}{ccccccc}
	Subs\footnote{This column denotes the number of subdesigns removed.} & $n$ & $\rk H$ & $\ k\ $ & $\ d\ $ & $\ c\ $ & Rate\\\hline
0 & 651 & 57 & 538 & 4 & 1 & 0.8264\\
1 & 644 & 57 & 532 & 4 & 2 & 0.8370\\
2 & 637 & 57 & 526 & 4 & 3 & 0.8477\\
3 & 630 & 57 & 520 & 4 & 4 & 0.8587\\
4 & 623 & 57 & 514 & 4 & 5 & 0.8700\\
5 & 616 & 57 & 508 & 4 & 6 & 0.8815\\
6 & 609 & 57 & 502 & 4 & 7 & 0.8933\\
7 & 602 & 57 & 496 & 4 & 8 & 0.9053\\
8 & 595 & 57 & 490 & 4 & 9 & 0.9176\\
9 & 588 & 57 & 482 & 4 & 8 & 0.9269
\end{tabular}
\end{ruledtabular}
\end{table}

\subsubsection{Block-by-point (Type I) Projective geometry codes}
Next we consider EAQECCs obtained via Theorem \ref{const3}
by using the block-by-point incidence matrix of $PG_{1}(m,q)$.
The codes obtained in this manner correspond to the classical Type I LDPC codes.
As in the classical setting, Type I entanglement-assisted quantum regular LDPC codes
can correct many quantum errors.
Because an incidence matrix of $PG_1(m,q)$ for $q$ odd is almost full rank,
the corresponding Type I code is not of much interest.
Hence, in this portion we always assume that $q = 2^t$ for some positive integer $t$.
\begin{theorem}\label{th:pgevenq1}
For every pair of integers $t \geq 1$ and $m \geq 2$ there exists an entanglement-assisted quantum \textup{LDPC} code with girth six whose parameters $[[n, k, d; c]]$ are
\[n = \frac{2^{t(m+1)}-1}{2^{t}-1},\]
\[k = \frac{2^{t(m+1)}-1}{2^{t}-1} - 2\varphi(m,2^t)+c,\]
\[d = (2^{t}+2)2^{t(m-2)},\mbox{ and}\]
\[c \leq \varphi(m,2^t).\]
\end{theorem}

\Proof
Let $H\tr$ be a block-by-point incidence matrix of $PG_{1}(m,2^t)$.
Then we have $\rk H\tr H \leq \rk H = \varphi(m,2^t)$, where $\varphi(m,2^t)$ 
is given by Theorem \ref{thm:hamada}.
By a result of Calkin, Key, and de Resmini \cite{CKR}, the minimum distance 
of the binary linear code with parity-check matrix $H^T$ is $(2^t+2)2^{t(m-2)}$.
Applying Proposition \ref{basicconst2} proves the assertion.\qed

Note that here the distance grows exponentially as the dimension of the geometry increases.
When $m =2$, the EAQECCs are based on projective planes.
As shown in Subsection \ref{principles}, the EAQECC obtained from a Desarguesian projective plane of order $2^t$
consumes only one initial ebit.
Basing on Hamada's conjecture, we expect that in general the EAQECCs given in Theorem \ref{th:pgevenq1} consume relatively small numbers of ebits.

It is not clear from the formula for $\varphi(m,2^t)$ whether the net rate
of a Type I EAQECC based on $PG_1(m,2^t)$ is positive.
In order to produce useful catalytic quantum codes,
it is important to understand when the net rate is positive.
\begin{proposition}\label{lem:pg2dimok}
Let $H$ be an incidence matrix of $PG_{1}(2,2^{t})$.
Then for all $t \geq 2$ the \textup{EAQECC} obtained from $H\tr$ has a positive net rate.
\end{proposition}
\Proof By Hamada's formula, we have $\rk H = 3^{t}+1$. The number of points in $PG_{1}(2,2^{t})$ is $2^{2t}+2^{t}+1$.\qed

For $m \geq 3$, we note that as $q$ increases,
$\rk H$ grows at a slower rate than the code length.
Thus we may expect that, for $q$ large when compared to $m$, the net rate will eventually become positive.
For example, one can check that the net rate of the Type I EAQECC obtained from $PG_1(3,2^t)$ is positive for $t \geq 7$.
Table \ref{fig:pg1params} gives sample parameters of the Type I codes obtained from $PG_{1}(m,2^t)$.
\begin{table}[h!t]\caption{Sample parameters of Type I $\qecci{n}{k}{d}{c}$ EAQECCs obtained from $PG_{1}(m,q)$, $q$ even.\label{fig:pg1params}}
\begin{ruledtabular}
\begin{tabular}{llllll}
$m$ & $q$ & $n$ & $k$ & $d$ & $c$ \\\hline
2 & 4 & 21 & 2 & 6 & 1\\
2 & 8 & 73 & 18 & 10 & 1\\
2 & 16 & 273 & 110 & 18 & 1\\
2 & 32 & 1057 & 570 & 34 & 1
\end{tabular}
\end{ruledtabular}
\end{table}

\subsection{Affine geometry codes\label{sub:ag}}
In this subsection, we will study the EAQECCs obtained from affine geometry designs.

The affine geometry $AG(m,q)$ of dimension $m$ over $\fq$ is a finite geometry whose points are the vectors in $\fq^{m}$.
The $i$-dimensional affine subspaces (or \emph{$i$-flats}) are the $i$-dimensional vector subspaces of $\fq^{m}$ and their cosets.
Thus $AG(m,q)$ has a natural parallelism.

The points and lines (that is, 1-flats) of an affine geometry form an $S(2,q,q^m)$, denoted by $AG_{1}(m,q)$.
The design has $q^{m-1}\frac{q^{m}-1}{q-1}$ blocks and replication number $\frac{q^{m}-1}{q-1} = q^{m-1}+q^{m-2}+ \cdots + q + 1$.

We note that in many papers concerning LDPC codes, the term ``Euclidean geometry'' and the notation $EG(m,q)$ are used for affine geometries.
Most of the codes studied in relation to Euclidean geometries does not use the zero vector, and hence they do not generally correspond to $S(2,\ka,v)$s.
Because the term ``affine geometry" is standard in the recent research on finite geometry in mathematics,
we use the notation $AG_1(m,q)$ when we take all points and lines to form an incidence matrix.
The incidence structure obtained by discarding the zero vector and the lines containing the zero vector from $AG_1(m,q)$ will be denoted by $EG_1(m,q)$,
which we will study in Subsection \ref{sub:eg}.
Because many of the classical FG-LDPC codes obtained from affine geometries are based on $EG_1(m,q)$,
they are generally not the same as the affine geometry codes presented in this section.

As with projective geometry designs, Propositions \ref{basicconst1} and \ref{basicconst2} give Type II and Type I affine geometry codes respectively.
It is notable that the classical ingredients of these codes are quasi-cyclic LDPC codes similar to other FG-LDPC codes
because the elementary abelian group acts transitively on the points of $AG_{1}(m,q)$ (see \cite{BJL,KLF}).
The rank of an affine geometry design $AG_{1}(m,2^{t})$ is directly related to $\varphi$ given in Theorem \ref{thm:hamada}.
\begin{theorem}[Hamada \cite{Ham73}]\label{thm:agpgrk} The rank of the affine geometry
design $AG_{1}(m,2^{t})$ is given by
\[\rk AG_{1}(m,2^{t}) = \varphi(m,2^t) - \varphi(m-1,2^t).\]
\end{theorem}
If $q$ is odd, the rank of $AG_1(m,q)$ over ${\mathbb{F}_{2}}$ is full.
\begin{theorem}[Yakir \cite{Y}]\label{thm:agoddrk} Let $H$ be an incidence matrix of the design $AG_{1}(m,q)$ with $q$ odd. Then $\rk H = q^{m}$.
\end{theorem}
Thus the dimensions of the corresponding FG-LDPC codes can be easily
calculated.

As in the case of projective designs, Hamada conjectured that the rank of $AG_{1}(m,2^{t})$ is minimum
among all Steiner $2$-designs of the same order and block size.
Thus, affine geometry designs with $q$ even may be expected to give codes with the best possible dimensions among all non-isomorphic $S(2,2^{t},2^{tm})$s.

We divide this subsection into two portions. In the first portion we examine high rate Type II entanglement-assisted quantum LDPC codes
obtained from $AG_{1}(m,q)$. Then in the next portion we present Type I entanglement-assisted quantum LDPC codes
based on $AG_{1}(m,q)$, which effectively exploit the redundancy to give excellent error correction performance.

\subsubsection{Point-by-block (Type II) Affine geometry codes}
The geometric structure of affine geometry has often been studied independently in various fields.
The special substructure we need to give distances has been investigated
in connection with the disk failure resilience ability of a class of redundant arrays of independent disks (RAID).
Here we present a known result on RAID related to our codes in coding theoretic terminology.
\begin{theorem}[M\"{u}ller and Jimbo \cite{MJ}]\label{jimbo}
Let $H$ be an incidence matrix of $AG_1(m,q)$.
The minimum distance of the classical binary linear code having $H$ as a parity-check matrix is $q+1$ if $q$ is even, and $2q$ otherwise.
\end{theorem}

The following two theorems give infinite families of EAQECCs which consume only one initial ebit and have very large net rate.
\begin{theorem}\label{th:agevenq2}
For every pair of integers $t \geq 1$ and $m \geq 2$
there exists an entanglement-assisted quantum \textup{LDPC} code with girth six whose parameters $[[n, k, d; c]]$ are
\[n = 2^{t(m-1)}\frac{2^{tm}-1}{2^{t}-1},\]
\[k = 2^{t(m-1)}\frac{2^{tm}-1}{2^{t}-1} - 2(\varphi(m,2^{t})-\varphi(m-1,2^{t})) + 1,\]
\[d = 2^{t}+1, \mbox{ and}\]
\[c = 1.\]
\end{theorem}

\Proof
Let $H$ be an incidence matrix of $AG_{1}(m,2^t)$.
By Theorem \ref{thm:agpgrk}, we have $\rk{H} = \varphi(m,2^{t}) - \varphi(m-1,2^{t})$.
The index of the design $AG_{1}(m,2^t)$ is one. Its replication number is always odd.
Thus, by Theorem \ref{const1}, we have $\rk{HH^T} = 1$.
Applying Proposition \ref{basicconst1} and Theorem \ref{jimbo} completes the proof.\qed
\begin{theorem}\label{th:agoddq2}
Let $q$ be an odd prime power.
Then for every integer $m \geq 2$ there exists an entanglement-assisted quantum \textup{LDPC} code with girth six whose parameters $[[n,k,d;c]]$ are
\[n = q^{m-1}\frac{q^{m}-1}{q-1},\]
\[k = q^{m-1}\frac{q^{m}-1}{q-1} - 2q^{m}+c,\]
\[d = 2q,  \ \mbox{and}\]
\[c =  \left \{\begin{array}{ccl}
1 & \mbox{when} & m \mbox{ is odd,}\\
q^{m}-1 & \mbox{when} & m \mbox{ is even.}
\end{array} \right.\]
\end{theorem}
\Proof Let $H$ be an incidence matrix of $AG_{1}(m,q)$ with $q$ odd.
By Theorem \ref{thm:agoddrk}, we have $\rk H = q^{m}$.
The index of the design $AG_{1}(m,q)$ is one. Its replication number $r$ is a sum of $m$ terms, each being an odd number.
Thus $r$ is odd only when $m$ is odd.
By Theorem \ref{const1}, we have $\rk{HH^T} = 1$ for $m$ odd. If $m$ is even, we have $\rk{HH^T} = q^{m}-1$ from Theorem \ref{const2}.
Applying Proposition \ref{basicconst1} and Theorem \ref{jimbo} proves the assertion.\qed

Theorem \ref{th:agoddq2} gives an infinite family of high rate entanglement-assisted quantum LDPC codes
which exploit reasonable amounts of entanglement as well.
Tables \ref{fig:ag2paramseven} and \ref{fig:ag2paramsodd} give a sample of the parameters of the Type II codes
obtained from $AG_{1}(m,q)$ with $q$ even and $q$ odd respectively.
\begin{table}[h!t]\caption{Sample parameters of Type II $\qecci{n}{k}{d}{c}$ EAQECCs obtained from $AG_{1}(m,q)$, $q$ even.\label{fig:ag2paramseven}}
\begin{ruledtabular}
\begin{tabular}{llllll}
$m$ & $q$ & $n$ & $k$ & $d$ & $c$ \\\hline
3 & 2 & 28 & 15 & 3 & 1\\
4 & 2 & 120 & 91 & 3 & 1\\
5 & 2 & 496 & 435 & 3 & 1\\
6 & 2 & 2016 & 1891 & 3 & 1\\
2 & 4 & 20 & 3 & 5 & 1\\
3 & 4 & 336 & 235 & 5 & 1\\
4 & 4 & 5440 & 4971 & 5 & 1\\
2 & 8 & 72 & 19 & 9 & 1\\
3 & 8 & 4672 & 3927 & 9 & 1
\end{tabular}
\end{ruledtabular}
\end{table}
\begin{table}[h!t]\caption{Sample parameters of Type II $\qecci{n}{k}{d}{c}$ EAQECCs obtained from $AG_{1}(m,q)$, $q$ odd.\label{fig:ag2paramsodd}}
\begin{ruledtabular}
\begin{tabular}{llllll}
$m$ & $q$ & $n$ & $k$ & $d$ & $c$ \\\hline
3 & 3 & 117 & 64 & 6 & 1\\
3 & 5 & 775 & 526 & 10 & 1 \\
3 & 7 & 2793 & 2108 & 14 & 1\\
5 & 3 & 9801 & 9316 & 6 & 1\\
4 & 3 & 1080 & 998 & 6 & 80
\end{tabular}
\end{ruledtabular}
\end{table}

Next we show that affine geometry designs have numerous subdesigns and Steiner spreads,
which make it possible to fine-tune the parameters and error correction performance of the corresponding EAQECCs.
\begin{theorem}\label{lem:affsubdesigns}
If $m \geq 3$, the points of $AG_{1}(m,q)$
can be partitioned into $q$ disjoint subsets of size $q^{m-1}$,
being the point sets of subdesigns isomorphic to $AG_{1}(m-1,q)$.
\end{theorem}
\Proof Take a parallel class $\{H_{1}, \ldots, H_{q}\}$ of $q$ hyperplanes
of $AG(m,q)$. Let the point set of $H_{j}$ be $V_{j}$.
Clearly $\cup_{j=1}^{q}V_{j} = V$, and the set of all blocks of $AG_{1}(m,q)$
which are contained entirely in $H_{j}$ form a subdesign isomorphic
to $AG_{1}(m-1,q)$.\qed

Theorem \ref{lem:affsubdesigns} can
be applied recursively to create
additional disjoint subdesigns of smaller dimension,
giving a variety of EAQECCs via Theorems \ref{ratecontrol}, \ref{ratecontrolspread}, and \ref{ratecontroloddr}.
Similar subdesign deletion techniques based on Theorem \ref{const2} further give infinitely many new high rate EAQECCs.
Table \ref{fig:ag134subdesigns} lists the parameters of the EAQECCs created by spread deletion from $AG_1(3,4)$.
\begin{table}[h!t]
\caption{Summary of Type II codes obtained by deleting a Steiner spread of subdesigns isomorphic to $AG_{1}(2,4)$ from $AG_{1}(3,4)$.\label{fig:ag134subdesigns}}
\begin{ruledtabular}
\begin{tabular}{ccccccc}
	Subs\footnote{This column denotes the number of subdesigns removed.} & $n$ & $\rk H$ & $\ k\ $ & $\ d\ $ & $\ c\ $ & Rate\\\hline
0 & 336 & 51 & 235 & 5 & 1 & 0.6994\\
1 & 316 & 51 & 216 & 5 & 2 & 0.7468\\
2 & 296 & 51 & 197 & 5 & 3 & 0.8007\\
3 & 276 & 51 & 178 & 5 & 4 & 0.8623\\
4 & 256 & 51 & 158 & 6 & 4 & 0.9297
\end{tabular}
\end{ruledtabular}
\end{table}

\subsubsection{Block-by-point (Type I) Affine geometry codes}
Next we consider the EAQECC obtained from a block-by-point incidence matrix of $AG_{1}(m,q)$.
Because incidence matrices of $AG_1(m,q)$ with $q$ odd are of full rank,
here we always assume $q = 2^t$ to obtain interesting codes.
The entanglement-assisted quantum LDPC codes presented in this section effectively exploit redundancy.
The excellent error correction performance will be demonstrated in simulations in Section \ref{sim}.

\begin{theorem}[Calkin, Key, and de Resmini \cite{CKR}]\label{th:distanceageven}
Let $H$ be a block-by-point incidence matrix of $AG_1(m,2^t)$.
Then the minimum distance of the classical binary linear code for which $H$ is a parity-check matrix is $(2^t+2)2^{t(m-2)}$.
\end{theorem}

\begin{theorem}\label{th:agevenq1}
For every pair of integers $t \geq 1$ and $m \geq 3$ there exists an entanglement-assisted quantum \textup{LDPC} code with girth six whose parameters $[[n, k, d; c]]$ are
\[n = 2^{tm},\]
\[k = 2^{tm}-2(\varphi(m,2^{t}) - \varphi(m-1,2^{t}))+c,\]
\[d = (2^{t}+2)2^{t(m-2)},\mbox{ and}\]
\[c \leq \varphi(m,2^{t}) - \varphi(m-1,2^{t}).\]
\end{theorem}

\Proof
Let $H\tr$ be a block-by-point incidence matrix of $AG_{1}(m,2^t)$.
By Theorem \ref{thm:agpgrk}, we have $\rk H\tr H \leq \rk H = \varphi(m,2^{t}) - \varphi(m-1,2^{t})$.
By Theorem \ref{th:distanceageven}, the minimum distance of the binary linear code with a parity-check matrix $H$ is $(2^t+2)2^{t(m-2)}$.
The assertion follows from Proposition \ref{basicconst2}.\qed

It is worth mentioning that here the distance grows exponentially with linear
increase of the geometry dimension $m$.
Because the rank of $AG_1(m,2^t)$ is conjectured to be the smallest possible
among all non-isomorphic $S(2,2^t,2^{tm})$s,
we expect that the EAQECCs obtained from these affine geometry designs consume the smallest possible numbers of ebits
attainable by this method with $S(2,2^t,2^{tm})$s.

When $m=2$, we can easily determine the required amount of entanglement.
\begin{theorem}\label{th:agevenq1m2}
For every positive integer $t$ there exists an entanglement-assisted quantum \textup{LDPC} code with girth six whose parameters $[[n, k, d; c]]$ are
\[n = 4^{t},\]
\[k = 4^{t}+2^{t}-2\cdot3^{t},\]
\[d = 2^{t}+2,\mbox{ and}\]
\[c = 2^{t}.\]
\end{theorem}
\Proof
Let $H\tr$ be a block-by-point incidence matrix of $AG_{1}(2,2^t)$.
We first prove that $\rk H\tr H = 2^t$.
Two lines of an affine plane are either parallel or intersect in exactly one point.
There are $2^t+1$ parallel classes of lines, each containing exactly $2^t$ lines, and each line contains $2^t$ points.
Because $2^t$ is even, it is always possible to reorder the rows of $H^T$ such that $H^TH$ is a block matrix of the following form:
\[H\tr H = \left[\begin{array}{cccc}
0 & J & & J \\
J & 0 & \cdots & J \\
& \vdots & \ddots & \vdots\\
J & J & \cdots & 0
\end{array}\right]
\]
where $J$ is the $2^t \times 2^t$ all-one matrix.
Hence, we have $\rk H\tr H = 2^t$.
By Theorem \ref{thm:agpgrk}, we have $\rk{H} = 3^t$.
Applying Proposition \ref{basicconst2} and Theorem \ref{th:distanceageven} 
completes the proof.\qed

Table \ref{fig:ag1paramseven} gives sample parameters of the Type I EAQECCs obtained from $AG_{1}(m,2^t)$.
\begin{table}[h!t]\caption{Sample parameters of Type I $\qecci{n}{k}{d}{c}$ EAQECCs obtained from $AG_{1}(m,q)$, $q$ even.\label{fig:ag1paramseven}}
\begin{ruledtabular}
\begin{tabular}{llllll}
$m$ & $q$ & $n$ & $k$ & $d$ & $c$ \\\hline
2 & 8 & 64 & 18 & 10 & 8\\
2 & 16 & 256 & 110 & 18 & 16\\
2 & 32 & 1024 & 570 & 34 & 32\\
2 & 64 & 4096 & 2702 & 66 & 64
\end{tabular}
\end{ruledtabular}
\end{table}

\subsection{Euclidean geometry codes\label{sub:eg}}
In this final subsection concerning finite geometry EAQECCs, we will examine Euclidean geometry codes.

Given a prime power $q$ and integer $m\ge 2$,
we define an incidence structure $EG_{1}(m,q)$ having as points
all points of $AG_{1}(m,q)$ except the zero vector, and having as blocks (or lines)
all lines of $AG(m,q)$ except those lines containing the zero vector.
The lines which are excluded from $AG_1(m,q)$ to form $EG_{1}(m,q)$ consist
of all multiples of a single nonzero vector. Thus, $EG_{1}(m,q)$ has $q^{m}-1$
points and $\left(q^{m-1}-1\right)\frac{q^{m}-1}{q-1}$ lines.
Each line contains $q$ points, and each point appears in
$\frac{q^{m}-1}{q-1} - 1 = q^{m-1} + q^{m-2} + \cdots + q$ lines.
Thus,  $EG_1(m,q)$ yields regular LDPC codes. Each pair of points appears in
\emph{at most} one line. Hence, $EG_{1}(m,q)$ is a partial Steiner 2-design.
Its Tanner graph does not contain 4-cycles.

Applying Proposition \ref{basicconst2} to a line-by-point incidence matrix of $EG_{1}(m,q)$
gives a Type I EAQECC.
If $q$ is even, the distance is bounded from below by the BCH bound.
\begin{theorem}[Kou, Lin, and Fossorier \cite{KLF}]\label{th:egdistance}
Let $H$ be a line-by-point incidence matrix of $EG_1(m, 2^t)$.
Then the minimum distance $d$ of the classical binary linear code having $H$ as a parity-check matrix satisfies
$d \geq \frac{2^{tm}-1}{2^t-1}$. Equality holds if $m=2$.
\end{theorem}

We use the following theorem to give the dimensions of FG-LDPC codes obtained from $EG_{1}(m,2^t)$
and their entanglement-assisted quantum counterparts.
\begin{theorem}[Hamada \cite{Ham73}]\label{thm:egpgrk}
The rank of the incidence structure $EG_{1}(m,2^{t})$, $t > 1$, is given by \[\rk EG_{1}(m,2^{t}) = \varphi(m,2^{t}) - \varphi(m-1,2^{t}) - 1.\]\end{theorem}

\begin{theorem}\label{th:egevenq2}
For every pair of integers $t \geq 1$ and $m \geq 2$  there exists an entanglement-assisted quantum \textup{LDPC} code with girth six whose parameters $[[n, k, d; c]]$ are
\[n = 2^{tm}-1,\]
\[k = 2^{tm}-2(\varphi(m,2^{t})-\varphi(m-1,2^{t}))+1+c,\]
\[d \geq \frac{2^{tm}-1}{2^{t}-1}, \mbox{ and}\]
\[c \leq \varphi(m,2^{t})-\varphi(m-1,2^{t})-1.\]
\end{theorem}
\Proof Let $H\tr$ be a line-by-point incidence matrix of $EG_{1}(m,2^t)$.
By Theorem \ref{thm:egpgrk}, we have $\rk H\tr H \leq \rk H = \varphi(m,2^{t})-\varphi(m-1,2^{t})-1$.
Applying Proposition \ref{basicconst2} and Theorem \ref{th:egdistance} completes the proof.\qed

A simple observation gives exact values of all the parameters of the Type I codes based on $EG_{1}(2,2^{t})$.
\begin{theorem}\label{cor:egevenq2m2}
For every positive integer $t$ there exists an entanglement-assisted quantum \textup{LDPC} code with girth six whose parameters $[[n, k, d; c]]$ are
\[n = 4^{t}-1,\]
\[k = 4^{t}+2^{t}-2\cdot 3^{t}+1,\]
\[d = 2^{t}+1, \mbox{ and}\]
\[c = 2^{t}.\]
\end{theorem}

\Proof
Let $H\tr$ be a line-by-point incidence matrix of $EG_{1}(2,2^t)$.
An incidence matrix of $EG_{1}(2,2^t)$ is obtained by removing one row and one column from each block from that of $AG_1(2,2^t)$.
By following the argument in Theorem \ref{th:agevenq1m2}, it is straightforward to see that $\rk{H^TH} = 2^t$.
By Theorem \ref{thm:egpgrk}, we have $\rk H = \varphi(m,2^{t})-\varphi(m-1,2^{t})-1 = 3^t-1$.
Theorem \ref{th:egdistance} and Proposition \ref{basicconst2} prove the assertion.\qed

Table \ref{fig:eg1paramseven} gives a sample of the parameters of the Type I codes obtained from $EG_{1}(2,2^t)$.
\begin{table}[h!t]\caption{Sample parameters of Type I $\qecci{n}{k}{d}{c}$ EAQECCs obtained from $EG_{1}(2,q)$, $q$ even.\label{fig:eg1paramseven}}
\begin{ruledtabular}
\begin{tabular}{llllll}
$m$ & $q$ & $n$ & $k$ & $d$ & $c$ \\\hline
2 & 8 & 63 & 19 & 9 & 8 \\
2 & 16 & 255 & 111 & 17 & 16\\
2 & 32 & 1023 & 571 & 33 & 32\\
\end{tabular}
\end{ruledtabular}
\end{table}

As with $S(2,\ka,v)$s, the incidence structure $EG_1(m,q)$ can also generate a high rate LDPC code with girth six.
Applying Proposition \ref{basicconst1} to incidence matrices,
we obtain Type II EAQECCs. Here we investigate their parameters.

\begin{theorem}\label{egtypeIIdistance}
The minimum distance of a Type II \textup{EAQECC} based on $EG_{1}(m,q)$ is $q+1$ if $q$ is even, and $2q$ if $q$ is odd and $m > 2$.
\end{theorem}
\Proof Consider any set of linearly dependent columns in an incidence matrix of $EG_1(m,q)$.
The same columns appear in the corresponding incidence matrix of $AG_1(m,q)$, but with a single zero coordinate added.
These columns are still dependent in $AG_1(m,q)$. 
Hence the minimum distance is upper bounded by Theorem \ref{jimbo}. Thus we need only to show lower bounds.

We begin with $q$ even. If $q = m = 2$, we can check by hand that the minimum distance is three.
Henceforth assume that $q > 2$ or $m > 2$.
Because the minimum distance of the code obtained from $AG_1(m,q)$ is $q+1$,
there exists a set $S$ of $q+1$ linearly dependent columns of an incidence matrix of $AG_1(m,q)$,
corresponding to a set ${\mathcal D}$ of $q+1$ blocks of $AG_{1}(m,q)$.
Let $P$ be the multiset of points appearing in the blocks of ${\mathcal D}$.
As each block of ${\mathcal D}$ has $q$ points, $|P| = q(q+1)$.
However, because the columns of $S$ are dependent over $\mathbb{F}_{2}$, each point in $P$ must appear with multiplicity two or more.
Hence, the number of distinct points in $P$ is at most $\frac{q(q+1)}{2} < q^{m}-1$ except for $q = m = 2$.
Therefore there is a nonzero point $p$ of $AG(m,q)$ which does not appear in $P$.
Let ${\mathcal D}' = \{B - p : B \in {\mathcal D}\}$, that is, we shift each block of ${\mathcal D}$ by $p$.
Each new block corresponds to a coset of a linear space.
Because $p \not\in P$, no element of ${\mathcal D}'$ contains the zero vector, and so the elements of ${\mathcal D}'$ are lines of $EG_{1}(m,q)$.
Thus ${\mathcal D}'$ is a linearly dependent set in $EG_{1}(m,q)$ of size $q+1$.
Therefore in all cases, the minimum distance of Type II EAQECC based on $EG_{1}(m,q)$, $q$ even, is $q+1$.
A similar argument proves the case when $q$ is odd and $m \neq 2$.\qed

\begin{theorem}\label{th:egsummary}
For every pair of integers $t \geq 1$ and $m \geq 2$  there exists an entanglement-assisted quantum \textup{LDPC} code with girth six whose parameters $[[n, k, d; c]]$ are
\[n = (2^{t(m-1)}-1)\frac{2^{tm}-1}{2^{t}-1},\]
\[k = (2^{t(m-1)}-1)\frac{2^{tm}-1}{2^{t}-1} - 2\rk{EG_{1}(m,2^{t})} + c,\]
\[d = 2^{t}+1, \mbox{ and}\]
\[c = \frac{2^{tm}-2^t}{2^{t}-1},\]
where $\rk{EG_{1}(m,2^{t})} = \varphi(m,2^{t}) - \varphi(m-1,2^{t}) - 1.$
\end{theorem}

\Proof
Let $H$ be an incidence matrix of $EG_1(m,2^t)$.
Because $H$ is obtained from an incidence matrix of $AG_1(m,2^t)$
by deleting the row representing the zero vector and the columns that represent the lines containing the zero vector,
it is easy to see that the rows and columns of $HH^T$ can be reordered such that the matrix is of the form:
\[HH^T = \left[\begin{array}{cccc}
0 & J & & J \\
J & 0 & \cdots & J \\
& \vdots & \ddots & \vdots\\
J & J & \cdots & 0
\end{array}\right]
\]
where $J$ is the $(2^t-1) \times (2^t-1)$ all-one matrix. Because $2^{tm}-1$ is odd, $\rk{HH^T}=\frac{2^{tm}-1}{2^{t}-1}-1$.
Applying Proposition \ref{basicconst1} and Theorems \ref{egtypeIIdistance} and \ref{thm:egpgrk} completes the proof.\qed

Tables \ref{fig:eg2paramseven} gives sample parameters for the Type II codes obtained from $EG_{1}(m,2^t)$.
\begin{table}[h!t]\caption{Sample parameters of Type II $\qecci{n}{k}{d}{c}$ EAQECCs obtained from $EG_{1}(m,q)$, $q$ even.\label{fig:eg2paramseven}}
\begin{ruledtabular}
\begin{tabular}{llllll}
$m$ & $q$ & $n$ & $k$ & $d$ & $c$ \\\hline
3 & 2 & 21 & 15 & 3 & 6 \\
4 & 2 & 105 & 91 & 3 & 14 \\
5 & 2 & 465 & 434 & 3 & 30 \\
6 & 2 & 1953 & 1891 & 3 & 62 \\
3 & 4 & 315 & 235 & 5 & 20 \\
4 & 4 & 5355 & 4971 & 5 & 84 \\
2 & 8 & 63 & 19 & 9 & 8 \\
3 & 8 & 4599 & 3927 & 9 & 72
\end{tabular}
\end{ruledtabular}
\end{table}

For the case $q$ odd,
Hamada \cite{Ham73} conjectured that an incidence matrix of $EG_1(m,q)$ is of full rank.
As shown in Table \ref{fig:eg2paramsodd}, the conjecture is true for small $m$ and $q$.
\begin{table}[h!t]\caption{Sample parameters of Type II $\qecci{n}{k}{d}{c}$ EAQECCs obtained from $EG_{1}(m,q)$, $q$ odd.\label{fig:eg2paramsodd}}
\begin{ruledtabular}
\begin{tabular}{llllll}
$m$ & $q$ & $n$ & $k$ & $d$ & $c$ \\\hline
3 & 3 & 104 & 64 & 6 & 12 \\
4 & 3 & 1040 & 960 & 6 & 80 \\
5 & 3 & 9680 & 9316 & 6 & 120 \\
3 & 5 & 744 & 526 & 10 & 30 \\
3 & 7 & 2736 & 2108 & 14 & 56
\end{tabular}
\end{ruledtabular}
\end{table}

\section{\label{sim}Performance}
In this section, we present simulation results for EAQECC codes constructed in the previous sections.
As in the related works \cite{HBD, HYH}, we performed simulations over the depolarizing channel.
In this model, each error ($X$, $Y$, and $Z$) occurs independently in each qubit with equal probability $f_{m}$.
For a given CSS type EAQECC,
we performed each decoding in two separate decoding steps, each using the sum-product algorithm.
The shared ebits, which do not pass through the noisy channel, are assumed to be error-free.
Our simulation results are reported in terms of the block error rate (BLER).

\begin{figure}[h!t]
\includegraphics[scale=0.53]{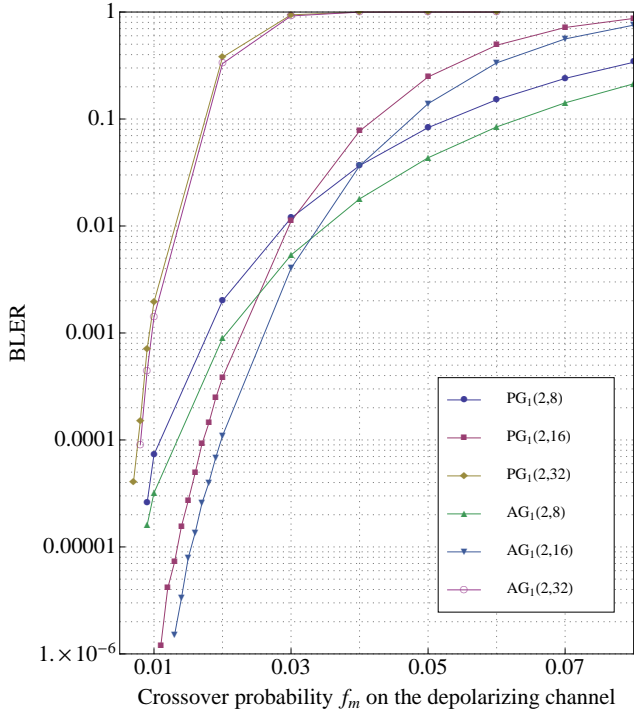}\caption{(Color online) Performance of Type I EAQECCs\label{fig:t1plot}}
\end{figure}
We first examine codes obtained from a block-by-point incidence matrix.
Figure \ref{fig:t1plot} shows the performance of several such codes based on projective and affine geometry designs.
As shown in Section \ref{finite}, these codes have very large distances for sparse-graph codes while avoiding short cycles.
As expected, these codes perform excellently at relatively high $f_m$.

To illustrate how well these codes perform,
we compare one of our Type I LDPC codes with previously known entanglement-assisted quantum LDPC codes with best BLERs.

Theorem \ref{th:agevenq1m2} gives a new EAQECC with parameters $[[256,110,18;16]]$ obtained from $AG_1(2,16)$.
The $[[255,111,17;16]]$ EAQECC in the work of Hsieh, Yen, and Hsu \cite{HYH} used $EG_1(2,16)$ outperformed
all previously known quantum codes of similar rate in simulations over the depolarizing channel.
Their code based on $PG_1(2,16)$, which also performed very well, has parameters $[[273,110,18;1]]$.
Exactly the same EAQECCs as these two can be constructed using Theorems \ref{cor:egevenq2m2} and \ref{plane} in our framework
without relying on computers to calculate their parameters.

\begin{figure}[h!t]
\includegraphics[scale=0.53]{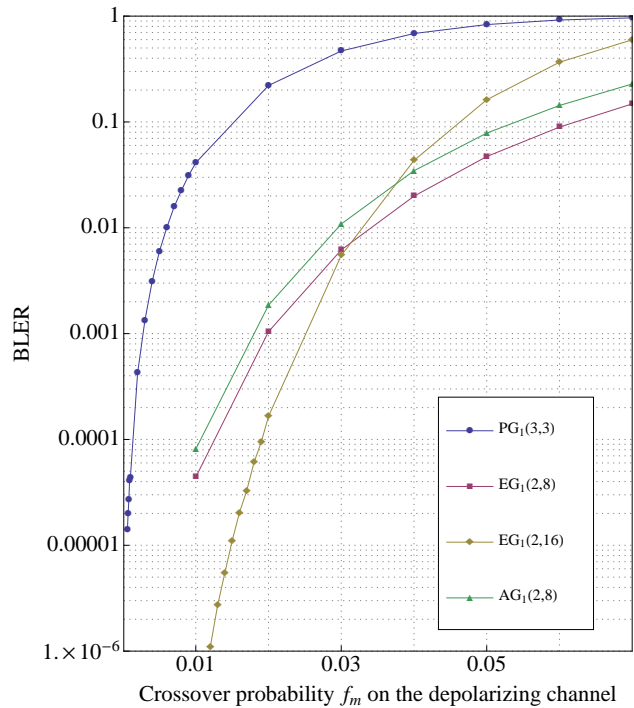}\caption{(Color online) Performance of Type II EAQECCs\label{fig:t2p1plot}}
\end{figure}

These three EAQECCs based on finite geometries have similar geometrical structures, and they behave quite similarly in simulations.
Performance of the $AG_1(2,16)$ and $PG_1(2,16)$ codes is directly compared in Figure \ref{fig:t1plot}.
The BLER of the $EG_1(2,16)$ code, which is slightly worse than that of our $AG_1(2,16)$ code, is plotted in Figure \ref{fig:t2p1plot}
to compare the three with EAQECCs having different parameters.
As shown in the figures, our new $[[256,110,18;16]]$ EAQECC obtained from $AG_1(2,16)$ shows a better BLER than the other two.
The BLERs of $AG_1(2,16)$, $EG_1(2,16)$, and $PG_1(2,16)$ codes at $f_m = 0.02$ are $1.0\times10^{-4}$, $1.6\times10^{-4}$, and $3.8\times10^{-4}$ respectively.

Entanglement-assisted quantum quasi-cyclic LDPC codes proposed by Hsieh, Brun, and Devetak in \cite{HBD} have also shown excellent BLERs.
In simulations their $[[128,58,6;18]]$ EAQECCs, called EX1 and EX2, outperformed the previously known best quantum LDPC codes at a similar rate about $0.316$.
The net rate of EX1 and EX2 is $\frac{58-16}{128} \approx 0.312$. 
Our $[[256,110,18;16]]$ EAQECC obtained from $AG_1(2,16)$ has net rate $\frac{110-16}{256} \approx 0.367$, which is higher than that of EX1 and EX2.
Their simulation results and our independent simulation results for EX1 and EX2 showed that their BLERs at $f_m = 0.02$ are higher than $1.1 \times 10^{-2}$
while our $AG_1(2,16)$ code has BLER about $1.0\times10^{-4}$ at the same $f_m$, which is better than EX1 and EX2 by two orders of magnitude.
Our EAQECC also requires a smaller amount of entanglement than EX1 and EX2.

Our results here confirm the close linkage between EAQECCs and classical error-correcting codes:
good performance in the classical setting translates directly into good performance from the corresponding quantum codes.

\begin{figure}[h!t]
\includegraphics[scale=0.53]{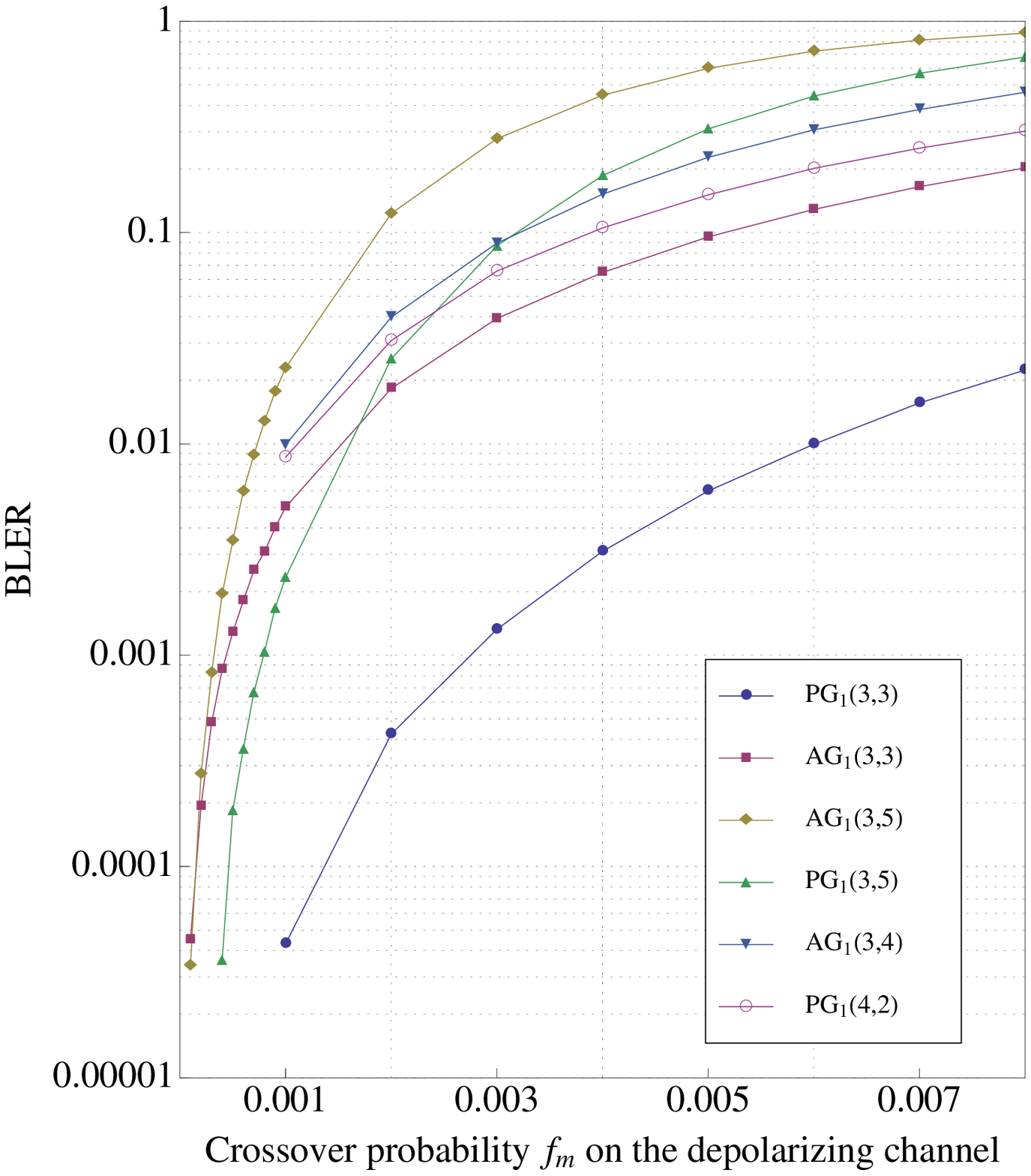}\caption{(Color online) Performance of Type II EAQECCs\label{fig:t2p2plot}}
\end{figure}
\begin{figure}[h!t]
\includegraphics[scale=0.53]{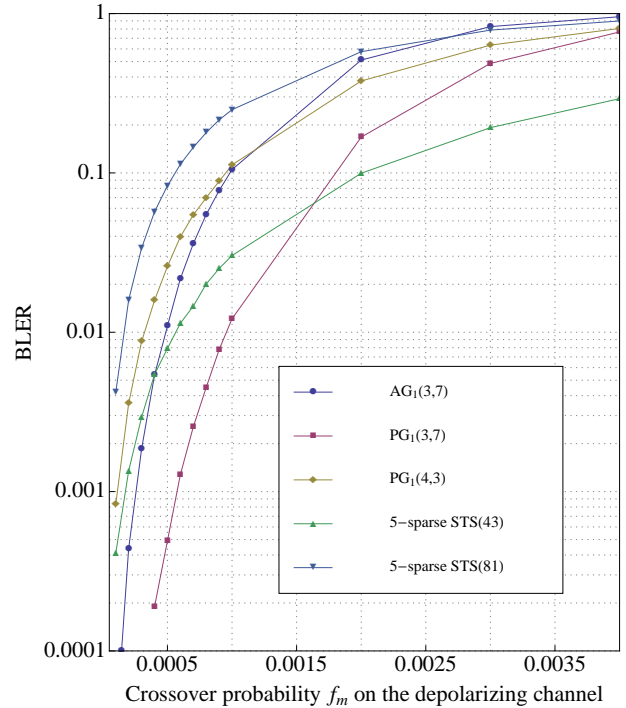}\caption{(Color online) Performance of high-rate Type II EAQECCs\label{fig:t2highrateplot}}
\end{figure}
We next examine codes obtained from a point-by-block incidence matrix.
These codes are capable of achieving extremely high rates even at moderate block lengths.

Figures \ref{fig:t2p1plot} and \ref{fig:t2p2plot} show the performance of several Type II codes based on finite geometries.
The Type II code from $PG_1(3,3)$ is shown in both figures to serve as a point of reference between the two figures.
Figure \ref{fig:t2highrateplot} gives the block error rates for several codes with high rates
including $[[301,216,6;1]]$ and $[[1080,998,6;80]]$ codes from cyclic $5$-sparse STSs of order $43$ and $81$ respectively.
The incidence matrices of these two Steiner triple systems are constructed from the list of base blocks in \cite{CMRS}.
Note that the cyclic automorphisms and sparse configurations immediately give the dimensions and distances of
the EAQECCs obtained from the cyclic $5$-sparse STSs (see \cite{DHV, sparseYF}).
Table \ref{fig:rates} lists the rates of selected finite geometry codes shown in figures.
\begin{table}[h!t]
\caption{Rates of EAQECCs obtained from finite geometries.\label{fig:rates}}
\begin{ruledtabular}
\begin{tabular}{ccccc}
Type & Geometry & $m$ & $q$ & Rate \\\hline
II & PG & 4 & 3 & $0.9008$\\
II & PG & 3 & 7 & $0.7203$\\
II & PG & 3 & 5 & $0.6166$\\
II & PG & 3 & 3 & $0.4076$\\
II & AG & 3 & 7 & $0.7547$\\
II & AG & 3 & 5 & $0.6787$\\
II & AG & 3 & 3 & $0.5470$\\
II & AG & 2 & 8 & $0.2638$\\
II & EG & 2 & 16 & $0.4352$\\
II & EG & 2 & 8 & $0.3015$\\
I & PG & 2 & 32 & $0.5392$\\
I & PG & 2 & 16 & $0.4029$\\
I & PG & 2 & 8 & $0.2465$\\
I & AG & 2 & 32 & $0.5566$\\
I & AG & 2 & 16 & $0.4296$\\
I & AG & 2 & 8 & $0.2812$
\end{tabular}
\end{ruledtabular}
\end{table}

As in the classical setting, our codes obtained from point-by-block incidence matrices
have waterfall regions at low $f_{m}$ and transmit at extremely high rates. 
This direct correlation in performance between the classical and quantum settings can also be seen
when codes require only one ebit.
It may be worth mentioning that changing geometries or choosing a non-geometric $S(2,\ka,v)$ can give
slightly different BLER curves.
It would be interesting to investigate theoretical methods for finding $S(2,\ka,v)$s with desirable performance curves in given situations.

\begin{figure}[h!t]
\includegraphics[scale=0.53]{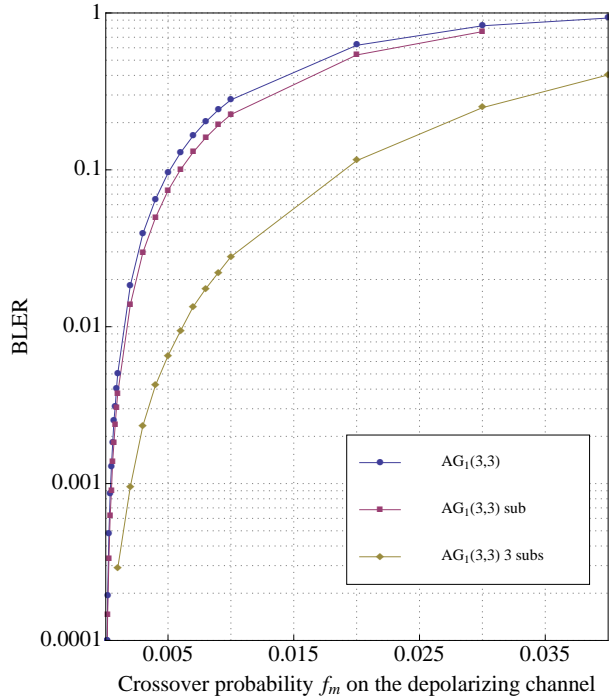}\caption{(Color online) Performance of EAQECCs obtained by deleting subdesigns from $AG_{1}(3,3)$.\label{fig:t2subdesigns}}
\end{figure}
Finally, we compare EAQECCs obtained by removing subdesigns from the parent design.
Here we test a subdesign deletion technique where each deletion step increases the required amount of entanglement
to a slightly larger degree than the examples we gave in Section \ref{finite}.
Each code in Figure \ref{fig:t2subdesigns} is constructed from a Type II code based on $AG_{1}(3,3)$.
Fundamental parameters of these codes are shown in Table \ref{fig:subdesigns}.
\begin{table}[h!t]
\caption{Summary of Type II EAQECCs obtained by deleting subdesigns from $AG_{1}(3,3)$.\label{fig:subdesigns}}
\begin{ruledtabular}
\begin{tabular}{ccccccc}
	Subs\footnote{This column denotes the number of subdesigns removed.} & $n$ & $\rk H$ & $\ k\ $ & $\ d\ $ & $\ c\ $ & Rate\\\hline
	0 & 117 & 27 & 64 & 6 & 1 & $0.5470$\\
	1 & 105 & 27 & 60 & 6 & 9 & $0.5714$\\
	2 & 93 & 26 & 58 & 6 & 17 & $0.6236$\\
	3 & 81 & 25 & 56 & 6 & 25 & $0.6913$
\end{tabular}
\end{ruledtabular}
\end{table}
The original code is also shown for reference.
The code labeled ``one sub'' has had a single subdesign isomorphic to $AG_{1}(2,3)$ removed.
The code labeled ``3 subs'' has had a Steiner spread removed.
This last code is a regular LDPC code.
As can be seen from their BLERs, removing subdesigns has improved the error correction performance
while increasing the rate and maintaining many of the essential properties.

Because removing subdesigns can increase the required amount of entanglement in a flexible manner,
one can generate an EAQECC which effectively exploits preexisting entanglement.
For example, a high net rate code consuming only one ebit can turn into a heavily entanglement-assisted code
to achieve better error correction performance at the same $f_m$.
As illustrated in Table \ref{fig:subdesigns},
a $[[117,64,6;1]]$ code with a regular parity-check matrix becomes a $[[81,56,6;25]]$ code
with a regular parity-check matrix through gradual steps.

One can also fine-tune parameters and improve error correction performance
while almost keeping the extremely low required amount of entanglement
by applying Theorems \ref{ratecontrol} and \ref{ratecontrolspread}.
As shown in Section \ref{finite}, all FG-LDPC codes found in \cite{HYH} can be constructed using our method.
The subdesign deletion techniques further give infinitely many new codes by fine-tuning their parameters and error correction performance.
In this sense, our method gives many kinds of new and known excellent EAQECCs in a single framework.

\section{\label{conclude}Conclusion}
We have developed a general framework for constructing entanglement-assisted quantum LDPC codes using combinatorial design theory.
Our constructions generate infinitely many new codes with various desirable properties
such as high error correction performance, high rates, and requiring only one initial entanglement bit.
Our methods are flexible and allow us to design EAQECCs with desirable properties while requiring prescribed amounts of entanglement.
All quantum codes constructed in this paper can be efficiently decoded through the sum-product algorithm.

We have introduced many new families of entanglement-assisted quantum LDPC codes based on combinatorial designs
as well as determined all fundamental parameters of the well-known families of LDPC codes based on finite geometries for most cases.
Because the entanglement-assisted stabilizer formalism bridges classical and quantum codes in a direct manner,
these results on entanglement-assisted quantum LDPC codes are useful both in quantum and classical coding theories.

Our framework encompasses many previously proposed excellent quantum LDPC codes as well.
In fact, our method can also be applied to quantum LDPC codes under the standard stabilizer formalism
by employing the ideas found in \cite{Aly, Djordjevic}.

We have focused on the fundamental classes of combinatorial designs.
However, other classes of incidence structures may provide interesting results as well.
For example, the entanglement-assisted quantum LDPC codes presented in \cite{HBD} can be seen as incidence structures generated
from the so-called difference matrices and their generalizations (see \cite{HandbookCD} for the definition and basic facts about difference matrices).
More general families of combinatorial designs can have nested structures or similar strong orthogonal relations between two incidence matrices.
This kind of structure can give asymmetric quantum codes (see \cite{IM, SKR}).
Structures in finite geometry we did not employ may also give interesting quantum LDPC codes as well as classical LDPC codes
(see, for example, \cite{Peter1, Peter2}).
Because LDPC codes and sparse incidence structures are equivalent, we expect that our methods may be further generalized
to encompass a wider range of both new and known quantum LDPC codes in future work.

\begin{acknowledgments}
Yuichiro Fujiwara acknowledges support from Japan Society for the Promotion of Science (JSPS) Grants-in-Aid for Scientific Research 20$\cdot$5897.
David Clark and Vladimir Tonchev acknowledge support by NSA Grant H98230-10-1-0177.
A substantial part of the research was done while Yuichiro Fujiwara was visiting the Department of Mathematical Sciences of Michigan Technological University.
He thanks the department for its hospitality.
The authors would like to thank the anonymous referee for his valuable comments and suggestions.
\end{acknowledgments}

\appendix
\section{Existence of $2$-designs}\label{appendix1}
Here we discuss the existence of $2$-designs to be applied to our constructions given in Subsection \ref{principles}.
The following is the well-known asymptotic existence theorem.

\begin{theorem}[Wilson \cite{W1, W2, W3}]\label{asymptotic}
The necessary conditions for the existence of a $2$-$(v,\ka,\lambda)$ design,
$\lambda(v-1) \equiv 0 \pmod{\ka -1}$ and $\lambda v(v-1) \equiv 0 \pmod{\ka(\ka-1)}$,
are also sufficient if $v > v_{\ka,\lambda}$, where $ v_{\ka,\lambda}$ is a constant
depending only on $\ka$ and $\lambda$.
\end{theorem}

For $\ka\in\{3,4,5\}$, necessary and sufficient conditions for the existence of an $S(2,\ka,v)$ are known.
\begin{theorem}[Kirkman \cite{Kirkman}]\label{STSex}
There exists an \textup{STS}$(v)$ if and only if $v \equiv 1,3 \pmod{6}$.
\end{theorem}
\begin{theorem}[Hanani \cite{HANANIK4}]\label{S24ex}
There exists an $S(2,4,v)$ if and only if $v \equiv 1,4 \pmod{12}$.
\end{theorem}
\begin{theorem}[Hanani \cite{HANANIK5}]\label{S25ex}
There exists an $S(2,5,v)$ if and only if $v \equiv 1,5 \pmod{20}$.
\end{theorem}

For $\ka \geq 6$, the necessary and sufficient conditions on $v$ for the existence of
an $S(2,\ka,v)$ are not known in general, although for small values of $\ka$ substantial results are known.
For a comprehensive table of known Steiner $2$-designs, see \cite{HandbookCD}.

Theorems \ref{asymptotic}, \ref{STSex}, \ref{S24ex}, and \ref{S25ex} were proved by constructive methods.
Hence, these existence results allow us to construct infinitely many explicit examples of entanglement-assisted quantum LDPC codes.
It is worth mentioning that many of the known proofs of these theorems employ the same construction technique we used in Theorem \ref{GDDspread}.
In fact, most $S(2,\ka,v)$s in the original proofs of these existence theorems have either Steiner spreads or nontrivial subdesigns.

Numerous other constructions for $2$-designs also give explicit examples of $S(2,\ka,v)$s for a wide range of parameters.
A detailed treatment of STS$(v)$s is available in \cite{TRIPLESYSTEMS}.
Various constructions for $S(2,\ka,v)$s for many values of $\ka$ are also given in \cite{Hall}.

\section{Parameters of quantum and classical FG-LDPC codes with girth six}\label{appendix2}
Here we give tables of parameters of LDPC codes with girth six based on finite geometries.
Table \ref{fig:quantumsummary} gives parameters of entanglement-assisted quantum LDPC codes obtained from
$PG_1(m,q)$, $AG_1(m,q)$, and $EG_1(m,q)$.
Parameters of the corresponding classical FG-LDPC codes are listed in Table \ref{fig:classicalsummary}.

\begin{table*}\caption{Parameters of entanglement-assisted quantum LDPC codes from finite geometries.\footnote{All codes are $[[n,k,d;c]]$ EAQECCs obtained from $PG_1(m,q)$, $AG_1(m,q)$, or $EG_1(m,q)$. We omit EAQECCs which are created by subdesign deletion techniques or do not have dimension greater than one. $\varphi(m,2^{t})$ is given by Theorem \ref{thm:hamada} in Subsection \ref{sub:pg}. $\varrho(m,2^t)$ is defined as $\varrho(m,2^t) = \varphi(m,2^{t})-\varphi(m-1,2^{t})$.}\label{fig:quantumsummary}}
\begin{ruledtabular}
\begin{tabular}{cccc|ccccc}
Geometry & Type\footnote{Type refers to the traditional classification of FG-LDPC codes: Type I uses a line-by-point incidence matrix, while Type II uses the transposed (i.e., point-by-line) incidence matrix.} & $m$ & $q$ & $n$ & $k$ & $d$ & $c$ & girth \\\hline

PG & II & any & $2^{t}$ & $\frac{(q^{m+1}-1)(q^{m}-1)}{(q^{2}-1)(q-1)}$ & $\frac{(q^{m+1}-1)(q^{m}-1)}{(q^{2}-1)(q-1)} - 2\varphi(m,2^{t}) + 1$ &$q+2$ & $1$ & 6\\

PG & II & odd & odd & $\frac{(q^{m+1}-1)(q^{m}-1)}{(q^{2}-1)(q-1)}$ & $\frac{(q^{m+1}-1)(q^{m}-1)}{(q^{2}-1)(q-1)} - 2\frac{q^{m+1}-q}{q-1}+1$ & $2(q+1)$ & $1$ & 6\\

PG & II & even & odd & $\frac{(q^{m+1}-1)(q^{m}-1)}{(q^{2}-1)(q-1)}$  & $\frac{(q^{m+1}-1)(q^{m}-1)}{(q^{2}-1)(q-1)} - \frac{q^{m+1}-q}{q-1}$ &  $2(q+1)$ & $\frac{q^{m+1}-q}{q-1}$ & 6\\

PG & I & 2 & $2^{t}$ & $q^{2}+q+1$  & $q^{2}+q - 2 \cdot 3^{t}$ & $q+2$ & 1 & 6\\

PG & I & any & $2^{t}$ & $\frac{q^{m+1}-1}{q-1}$ & $\frac{q^{m+1}-1}{q-1}-2\varphi(m,2^{t})+c$ & $(q+2)q^{m-2}$ & $\leq \varphi(m,2^{t})$ & 6\\

AG & II & any & $2^{t}$ & $q^{m-1}\frac{q^{m}-1}{q-1}$ & $q^{m-1}\frac{q^{m}-1}{q-1} - 2\varrho(m,2^t) + 1$ & $q+1$ & 1 & 6\\

AG & II & odd & odd & $q^{m-1}\frac{q^{m}-1}{q-1}$ & $q^{m-1}\frac{q^{m}-1}{q-1} - 2q^{m} + 1$ & $2q$ & 1 & 6\\

AG & II & even & odd & $q^{m-1}\frac{q^{m}-1}{q-1}$ & $q^{m-1}\frac{q^{m}-1}{q-1} - q^{m} -1$ & $2q$ & $q^{m}-1$ & 6\\

AG & I & 2 & $2^{t}$ & $q^{2}$ & $q^{2} + q - 2\cdot3^{t}$ & $q+2$ & $q$ & 6\\

AG & I & any & $2^{t}$ & $q^{m-1}\frac{q^{m}-1}{q-1}$ & $q^{m-1}\frac{q^{m}-1}{q-1} - 2\varrho(m,2^t)+c$ & $(q+2)q^{m-2}$ & $\leq\varrho(m,2^t)$ & 6\\

EG & I, II\footnote{The codes obtained from either orientation of the incidence matrix are identical \cite{KLF}.}& 2 & $2^{t}$ & $q^{2}-1$ & $q^{2} + q - 2\cdot3^{t} + 1$ & $q+1$ & $q$ & 6\\

EG & II & any & $2^{t}$ & $\frac{(q^{m-1}-1)(q^{m}-1)}{q-1}$ & $\frac{(q^{m-1}-1)(q^{m}-1)}{q-1}-2\varrho(m,2^t)+2+c$ & $q+1$ & $\frac{q^m-q}{q-1}$ & 6
\end{tabular}
\end{ruledtabular}
\end{table*}
\begin{table*}\caption{Parameters of classical FG-LDPC codes.\footnote{We omit the cases when codes are created by subdesign deletion techniques or do not have enough dimension.}\label{fig:classicalsummary}}
\begin{ruledtabular}
\begin{tabular}{cccc|cccc}
Geometry &Type & $m$ & $q$ & $n$ & $k$ & $d$ & girth \\\hline

PG & II & any & $2^{t}$ & $\frac{(q^{m+1}-1)(q^{m}-1)}{(q^{2}-1)(q-1)}$ & $\frac{(q^{m+1}-1)(q^{m}-1)}{(q^{2}-1)(q-1)} - \varphi(m,2^{t})$ &$q+2$ & 6 \\

PG & II & any & odd & $\frac{(q^{m+1}-1)(q^{m}-1)}{(q^{2}-1)(q-1)}$  & $\frac{(q^{m+1}-1)(q^{m}-1)}{(q^{2}-1)(q-1)} - \frac{q^{m+1}-q}{q-1}$ &  $2(q+1)$ & 6\\

PG & I & any & $2^{t}$ & $\frac{q^{m+1}-1}{q-1}$ & $\frac{q^{m+1}-1}{q-1}-\varphi(m,2^{t})$ & $(q+2)q^{m-2}$ & 6\\

AG & II & any & $2^{t}$ & $q^{m-1}\frac{q^{m}-1}{q-1}$ & $q^{m-1}\frac{q^{m}-1}{q-1} - \varphi(m,2^{t})+\varphi(m-1,2^{t})$ & $q+1$ & 6 \\

AG & II & any & odd & $q^{m-1}\frac{q^{m}-1}{q-1}$ & $q^{m-1}\frac{q^{m}-1}{q-1} - q^{m}$ & $2q$ & 6 \\

AG & I & any & $2^{t}$ & $q^{m}$ & $q^{m} - \varphi(m,2^{t})+\varphi(m-1,2^{t})$ & $(q+2)q^{m-2}$ & 6\\

EG & I, II\footnote{The codes obtained from either orientation of the incidence matrix are identical \cite{KLF}.} & 2 & $2^{t}$  & $q^{2}-1$ & $q^{2} - 3^{t}$ & $q+1$ & 6\\

EG & II & any & $2^{t}$ & $(q^{m-1}-1)\frac{q^{m}-1}{q-1}$ & $(q^{m-1}-1)\frac{q^{m}-1}{q-1}-\varphi(m,2^{t}) + \varphi(m-1,2^{t})+1$ & $q+1$ & 6\\

EG & II & $\geq 3$ & odd & $(q^{m-1}-1)\frac{q^{m}-1}{q-1}$ & $\geq (q^{m-1}-1)\frac{q^{m}-1}{q-1}-q^m+1$\footnote{If Hamada's conjecture on the rank of $EG_1(m,q)$ \cite{Ham73} is true, the equation holds.} & $2q$ & 6
\end{tabular}
\end{ruledtabular}
\end{table*}

\end{document}